\documentclass[12pt]{article}
\usepackage{amsmath,amssymb,epsfig,amsfonts}
\usepackage{graphicx,subfigure}
\usepackage[usenames, dvipsnames]{color}
\usepackage{tikz}
\usepackage{hyperref}
\usepackage[nosort]{cite}
\usepackage{lscape}
\usepackage{rotating}

\usepackage{extarrows}

\usepackage{verbatim} 

\usepackage[top=1.25in,bottom=1.25in,right=0.85in,left=0.85in]{geometry}

\newcommand{\KM}{Donaldson matching}

\makeatletter

\newcommand{\ov}[1]{\overline #1}

\newcommand{\bZ}{\mathbb{Z}}
\newcommand{\bC}{\mathbb{C}}
\newcommand{\bP}{\mathbb{P}}
\newcommand{\bR}{\mathbb{R}}

\newcommand{\cN}{\mathcal{N}}

\newcommand{\cO}{\mathcal{O}}

\def\unit{{1\kern-.65ex {\rm l}}}
\def\1{{1\kern-.65ex {\rm l}}}

\newcommand{\sskip}{\vspace{.2cm}}

\newcount\hour \newcount\minute
\hour=\time \divide \hour by 60
\minute=\time
\def\now{%
\ifnum \hour<13
  \ifnum \hour=0 \advance \hour by 12 \number\hour:\else \number\hour:\fi%
     \ifnum \minute<10 0\fi%
     \number\minute%
\ A.M.%
\else \advance \hour by -12 \number\hour:%
  \ifnum \minute<10 0\fi%
  \number\minute%
  \ P.M.%
\fi%
}

\makeatother

\begin{document}

\baselineskip=18pt  
\numberwithin{equation}{section}  
\allowdisplaybreaks  


%
%


\thispagestyle{empty}

\vspace*{-3cm}
\begin{flushright}
{\tt NSF-KITP-14-203}
\end{flushright}

\vspace*{1.8cm}
\begin{center} {\Huge The Landscape of M-theory Compactifications \\
    \vspace{.2cm} on Seven-Manifolds with $G_2$ Holonomy}

\vspace*{1.5cm}
James Halverson$^1$ and David R. Morrison$^2$
\vspace*{1.0cm}

$^1$ {\it Kavli Institute for Theoretical Physics, \\
University of California, Santa Barbara, CA 93106, USA}\\
{\tt jim@kitp.ucsb.edu} \\ \vspace{.5cm}

$^2$ {\it Departments of Mathematics and Physics, \\
University of California, Santa Barbara, CA 93106, USA}\\
{\tt drm@physics.ucsb.edu}

\vspace*{0.8cm}
\end{center}
\vspace*{1cm}

We study the physics of globally consistent four-dimensional $\mathcal{N}=1$
supersymmetric M-theory compactifications on $G_2$ manifolds
constructed via twisted connected sum; there are now perhaps
fifty million examples of these manifolds.  We study a rich example
that exhibits $U(1)^3$ gauge symmetry and a spectrum of massive
charged particles that includes a trifundamental. Applying recent
mathematical results to this example, we compute the form of
membrane instanton
corrections to the superpotential and spacetime topology change
in a compact model; the latter include both the (non-isolated) $G_2$ flop
and conifold transitions. The conifold transition spontaneously breaks
the gauge symmetry to $U(1)^2$, and associated field theoretic
computations of particle charges make correct predictions for the
topology of the deformed $G_2$ manifold. We discuss physical aspects of
the abelian $G_2$ landscape broadly, including aspects of Higgs and
Coulomb branches, membrane instanton corrections, and some general
aspects of topology change.


\clearpage
\tableofcontents

\newpage

\section{Introduction}

The landscape of four-dimensional string compactifications with
$\cN=1$ supersymmetry is vast. There are a variety of corners of the
landscape, and while certain special corners are well-controlled and amenable to detailed
calculations, it is often true that much less can be said about physics
in the broader regions.  This can be true for a number of reasons:
\begin{itemize}
\item[(1)] The theory could be strongly coupled.
\item[(2)] The theory could be at small volume.
\item[(3)] The relevant mathematical tools might not be adequately developed.
\end{itemize}
Of course, the extent to which these are drawbacks for an
understanding of any particular region of the landscape is
time-dependent.  However, it is sometimes the case that there are
techniques that allow for control of theories at strong coupling or
small volume that coincide with available mathematics.  For example,
the existence of construction techniques and knowledge of the relevant
moduli spaces in the case of Calabi-Yau varieties allows for the study
of many aspects of F-theory, despite the fact that the theory is
inherently strongly coupled.

Four-dimensional $\cN=1$ compactifications of M-theory with
non-abelian gauge symmetry are faced with all of these issues: (1) the
theory is a strongly coupled limit of the type IIa superstring, (2) the
existence of non-abelian gauge symmetry requires taking a singular
limit of the compactification manifold so that there is no large-volume
approximation, and (3) relatively little is known
about the relevant seven-manifolds (i.e., manifolds with $G_2$ holonomy)
compared to, for example,
Calabi-Yau threefolds.
Though the list of such seven-manifolds
has historically been rather sparse, the situation has improved in
recent
years due to the Kovalev
``twisted connected sum'' (TCS) construction
 \cite{KovalevTCS} which has been generalized and corrected
in recent years \cite{KovalevLee,MR3109862,arXiv:1212.6929,Corti:2012kd}.
The list of TCS examples is now large enough
to warrant speaking of a landscape of four-dimensional $\cN=1$ M-theory
compactifications on seven-manifolds with $G_2$ holonomy, which we will
refer to as the ``abelian $G_2$
landscape'' since these are compactifications on smooth manifolds and
hence have no non-abelian gauge symmetry.

\sskip \sskip How large is the abelian $G_2$ landscape? Saying something
quantitative requires being more specific about what one
means. Drawing a sharp analogy to type IIb vacua, one could mean the
number of de Sitter vacua and its dependence on the choice of M-theory
flux. However, making this analogy reliably within the
supergravity approximation would require restricting attention to vacua which do
not exhibit non-abelian gauge symmetry. Moreover, even in that
approximation the one-instanton effects from wrapped membranes likely play
an important role in moduli stabilization, and though we will discuss
progress in this direction it is not yet possible to say whether the
known instanton corrections are the leading instanton corrections. The
comparison to type IIb flux vacua is further complicated by the fact
that a classical flux superpotential may not play as significant of a
role in M-theory, since all geometric moduli may in principle be
stabilized by non-perturbative effects.

Perhaps a coarser comparison is more appropriate to estimate the size
of the abelian $G_2$ landscape: how many suitable compactification
manifolds exist, and how does this compare to the number of analogous
manifolds used for type IIb compactifications? This is really where
the recent gains \cite{KovalevTCS,KovalevLee,Corti:2012kd} have been
made. Since it is very familiar to physicists, it is worth comparing
to the Kreuzer-Skarke classification \cite{Kreuzer:2000xy} of
Calabi-Yau threefold hypersurfaces in certain four-dimensional toric
varieties.  In this case, there are four-dimensional toric varieties
associated to any reflexive polytope via its triangulations, and there
are 473,800,776 such polytopes. However, the Calabi-Yau hypersurfaces
associated to this list exhibit only 30,108 distinct Hodge pairs
$(h^{1,1},h^{2,1})$; though there is other topological data that may
distinguish the Calabi-Yau hypersurfaces, many of them may be
different realizations of the same Calabi-Yau. The heuristic lesson is
that the many different Calabi-Yau ``building blocks'' do not
necessarily give rise to distinct Calabi-Yau manifolds. By comparison,
the TCS construction of $G_2$ manifolds uses a pair of suitable
``building blocks'' and if there is a ``matching pair'' of building
blocks then a TCS $G_2$ manifold can be constructed from them, though
$G_2$ manifolds constructed from the same matching pair may be
topologically equivalent. In fact there are now
\cite{MR3109862,Corti:2012kd} at least fifty million such matching
pairs, and it stands to reason that the abelian $G_2$ landscape is now
quite large.

Aside from the ``landscape'' implied by a large number of example compactifications,
there is also evidence for a stronger notion of the word, as some topology
changing transitions between branches of $G_2$ moduli space are known, and we will
study the physics of one such example. Given these two facts, it is natural
to wonder whether a version of Reid's fantasy \cite{MR909231} for Calabi-Yau threefolds
also holds for $G_2$ manifolds; perhaps the many $G_2$ moduli spaces now known by the TCS
construction are part of one large connected irreducible $G_2$ landscape.

\sskip \sskip The purpose of this paper, which is complementary to our
work \cite{SingularLimits} on singular limits of $G_2$
compactifications, is to introduce the TCS construction into the
physics literature, study a rich example in detail, and to discuss
what can be said broadly about the physics of the abelian $G_2$
landscape using currently available mathematical results.  In section
\ref{sec:TCS review} we will review $G_2$ manifolds, $G_2$-structures,
and the TCS construction. We encourage the reader to read the following outline
carefully, since it also serves as a
summary of our results.

In section \ref{sec:example} we will study a TCS $G_2$
compactification and three branches of its moduli space. This globally
consistent compact model exhibits abelian gauge symmetry and massive
charged particles, a limit in moduli in which some particles become massless,
non-perturbative instanton corrections to the superpotential,
spontaneous symmetry breaking, and spacetime topology change via a non-isolated
flop or conifold transition.

We begin in section \ref{sec:relevant building block} by introducing a
TCS $G_2$ manifold studied in \cite{Corti:2012kd} that we call $X$,
focusing on one of the building blocks of that manifold; since
$b_2(X)=3$, the gauge symmetry of M-theory compactified on $X$ is $U(1)^3$. We
review the construction of that building block presented in
\cite{MR3109862}, but perform new computations of topological
intersections. We show that these intersections in the building block determine
two-cycle and five-cycle intersections in $X$, which in turn determine
the charges of massive particles arising from M2-branes wrapped on
two-cycles.

There are $24$ different massive charged particles, since there are
$24$ rigid holomorphic curves in the building block that become
two-cycles in $X$. In section \ref{sec:massive charged particles on X}
we compute their charges, which happen to include a trifundamental. By
a result of \cite{Corti:2012kd}, to each of these rigid holomorphic
curves in the building block there is an associated rigid associative
submanifold in $X$ diffeomorphic to $S^2\times S^1$; an M2-brane
instanton wrapped on such a cycle is expected to generate a
non-perturbative correction to the superpotential
\cite{Harvey:1999as}. We explicitly compute the form of the
non-perturbative superpotential in our example, which happens to be
intricate. There are $24$ rigid associatives in six different homology
classes with four representatives each. These generate a six term
non-perturbative superpotential, each with a prefactor $4$ that is the
$G_2$ analog of a Gromov-Witten invariant, and the total
non-perturbative superpotential depends on three moduli fields. This
superpotential is a generalized racetrack.

In section \ref{sec:massless limits, topology change, and the Higgs
  mechanism} we study limits in $G_2$ moduli space in which some of
the charged particles become massless. This is achieved using
calibrated geometry and a specific property of the calibrated
three-cycles, namely, that they contain non-trivial two-cycles. The
limit shrinks four three-cycles and two-cycles of the same respective
classes to zero size, yielding four circles of conifold points. We call this $G_2$
limit $X_c$, and M-theory compactified on $X_c$ has $U(1)^3$ gauge symmetry with
instanton corrections and four massless particles with the same
charge, as well as a spectrum of massive particles. Relative to $X_c$,
$X$ was a small resolution of the four circles of conifold points; the other small
resolution gives a $G_2$ manifold $X_s$ related to $X$ via a (non-isolated) flop transition,
and M-theory on $X$ and $X_s$ have similar physics. The circles of conifold points
can also be smoothed by a deformation within $G_2$ moduli to give a $G_2$ manifold $X_d$, where
M-theory on $X_d$ has $U(1)^2$ gauge symmetry. The gauge symmetry has been
spontaneously broken, and a simple field theoretic prediction related to
the charges of the Higgs fields correctly predicts aspects of the topology
of $X_d$, and the associated spectrum of massive particles for M-theory
on $X_d$.
This constructions are analogous to the original flop
\cite{Witten:1993yc,mult, catp} and conifold \cite{bhole} transitions
for Calabi--Yau threefolds, as well as their recent extension to
transitions for Calabi--Yau fourfolds \cite{4d-transitions}.\footnote{Note that our transitions are different
from the $G_2$ flop \cite{Atiyah:2000zz} and $G_2$ conifold \cite{Atiyah:2001qf}
transitions
which have previously been studied in a non-compact setting, since those involved isolated singularities.}

In section \ref{sec:landscape} we broadly discuss the physics of the
abelian $G_2$ landscape as determined by the topology of the known TCS
$G_2$ manifolds. For technical reasons, nearly all of the known
examples have $b_2=0$ and so do not exhibit even
abelian gauge symmetry, and therefore they must be
Higgs branches from the field theory viewpoint if singular limits with non-abelian gauge symmetry
exist for them. As we will have already shown in section 3, however, examples with abelian gauge symmetry do exist and may
arise in one of a few different ways that we review.  We discuss
membrane instanton corrections to the superpotential and potential
implications for moduli stabilization; a number of known examples
exhibit more than $40$ such corrections, and generalized racetracks
are to be expected (as seen in section 3). In \cite{Corti:2012kd} a number of general
statements are made about the topology of possible $G_2$ transitions;
we comment on the associated physical implications. We also discuss
some common model-building assumptions in light of the existence of
these new vacua.

In section \ref{sec:conclusions} we conclude, briefly discussing
needed mathematical progress that would be physically useful, as well
as future physical prospects, including for abelian de Sitter vacua.

\section{$G_2$ Manifolds from Twisted Connected Sums}
\label{sec:TCS review}
In this section we review Kovalev's construction \cite{KovalevTCS} for
obtaining compact $G_2$ manifolds from twisted connected
sums. We will review those results that would not be prudent
to review in the middle of the physics discussions of sections
\ref{sec:example} or \ref{sec:landscape} , for reasons of length or relevance.

The basic construction glues appropriate matching pairs of ``building
blocks'', each comprised of an algebraic threefold times a circle, to
give a $G_2$ manifold. In Kovalev's original work,
the building blocks were constructed from
Fano threefolds having a K3 surface in their
anticanonical class. Recently it has been shown \cite{Corti:2012kd} by
Corti, Haskins, Nordstr\"om, and Pacini (CHNP) that weak-Fano
threefolds --- which require only $-K \cdot C \geq 0 $ for $K$ the
canonical class and $C$ any holomorphic curve rather than the strict
inequality characteristic of Fano threefolds --- can also serve as
appropriate building blocks.  While a seemingly small change, this
adapted construction increases the number of matching pairs (and thus
$G_2$ manifolds) by orders of magnitude, from hundreds or thousands to
tens of millions.

For further details on the content of this section, we refer the
reader to \cite{Corti:2012kd}.

\subsection*{General Aspects of $G_2$ Manifolds}

Before presenting the TCS construction,
let us review some basic facts about $G_2$ manifolds in general, as
well as some conventions we will use throughout.

 A \emph{$G_2$-structure} on a seven-manifold $X$ is a principal
  subbundle of the frame bundle of $X$ that has structure group
  $G_2$. Practically, each $G_2$ structure is characterized by a three-form
  $\Phi$ and a metric $g_\Phi$ such that every tangent space of $X$ admits
  an isomorphism with $\bR^7$ that identifies $g_\Phi$ with $g_0\equiv
  dx_1^2 + \dots + dx_7^2$ and $\Phi$ with
  \begin{equation} \Phi_0 =
  dx_{123} + dx_{145} + dx_{167} + dx_{246} - dx_{257} - dx_{347} -
  dx_{356},
  \end{equation}
  where $dx_{ijk} \equiv dx_i \wedge dx_j \wedge dx_k$.  Note that the
  subgroup of $GL(7,\bR)$ which preserves $\Phi_0$ is the
  exceptional Lie group $G_2$ \cite{MR916718}. The three-form $\Phi$, sometimes called
  the \emph{$G_2$-form}, determines an orientation, the Riemannian
  metric $g_\Phi$, and a Hodge star $\star_\Phi$
  which we will often shorten to $\star$. We will refer to the pair
  $(\Phi,g_\Phi)$ as a $G_2$-structure.

  For a seven-manifold $X$ with a $G_2$-structure $(\Phi,g_\Phi)$
  and associated Levi-Civita connection $\nabla$, the \emph{torsion}
  of the $G_2$-structure is $\nabla \Phi$, and when $\nabla
  \Phi=0$ the $G_2$ structure is said to be
  \emph{torsion-free}. The following are equivalent:
  \begin{itemize}
  \item $Hol(g_\Phi) \subseteq G_2$
  \item $\nabla \Phi = 0$, and
  \item $d\Phi = d \star \Phi = 0$.
  \end{itemize}
  The triple $(X,\Phi,g_\Phi)$ is called a \emph{$G_2$-manifold} if
  $(\Phi,g_\Phi)$ is a torsion-free $G_2$-structure on $X$. Then by
  the above equivalence, the metric $g_\Phi$ has $Hol(g_\Phi)\subseteq
  G_2$ and $g_\Phi$ is Ricci-flat. For a compact $G_2$-manifold $X$,
  $Hol(g_\Phi)=G_2$ if and only if $\pi_1(X)$ is finite \cite{joyce2000compact}. In this case the
  moduli space of metrics with holonomy $G_2$ is a smooth manifold of
  dimension $b_3(X)$.

Calibrated geometry will be important in our work. In the absence of
explicit metric knowledge, as is typically the case for compact
Calabi-Yau or $G_2$ manifolds, the volumes of certain cycles can
nevertheless be computed via calibrated geometry as developed in the
seminal work of Harvey and Lawson \cite{HarveyLawson}. Their
fundamental observation is the following. Let $X$ be a Riemannian
manifold and $\alpha$ a closed $p$-form such that $\alpha|_\xi \leq
vol_\xi$ for all oriented tangent $p$-planes $\xi$ on $X$. Then any
compact oriented $p$-dimensional submanifold $T$ of $X$ with the
property that $\alpha|_T=vol_T$ is a minimum volume representative of
its homology class, that is
\begin{equation}
vol(T) = \int_T \alpha = \int_{T'} \alpha \leq vol(T')
\end{equation}
for any $T'$ such that $[T-T']=0$ in $H_p(X,\bR)$. Note in particular
the useful fact that $vol(T)$ is computed precisely by $\int_T
\alpha$, even though one may not know the metric on $X$.

If $X$ is a Calabi-Yau threefold, the K\" ahler form $\omega$ and the
holomorphic three-form $\Omega$ are calibration forms for two-cycles
and three-cycles; they calibrate holomorphic curves and special
Lagrangian submanifolds. Note, therefore, in M-theory compactifications
on $X$ the presence of calibrated two-cycles allows for control over
massive charged particle states obtained from wrapped M2-branes. This
computes particle masses as a function of moduli.

If $X$ is a $G_2$ manifold, $\Phi$ and $\star \Phi$ are calibration forms
which calibrate so-called associative three-cycles and coassociative
four-cycles, respectively. This allows for control over
topological defects obtained from wrapping M2-branes and M5-branes on
calibrated  three-cycles and four-cycles; these are instantons, domain
walls, and strings. Note the absence of calibrated two-cycles, however.

\subsection*{$G_2$ Structures on Product Manifolds}

Let $V$ be a K\"ahler manifold of complex dimension $3$ with K\"ahler form
$\omega$, and suppose $V$ has a nowhere-vanishing holomorphic $3$-form $\Omega$
satisfying the basic Calabi--Yau condition that $\Omega\wedge \overline{\Omega}$
is a constant times the volume form $\frac1{3!}\omega\wedge\omega\wedge\omega$.
(Notice that we are not insisting that $V$ be compact.)
Multiplying $\Omega$ by a suitable real constant if necessary, we may
assume that
\begin{equation} \label{eq:Omega normalization}
\frac i8\Omega \wedge \overline{\Omega} = \frac1{3!} \omega\wedge\omega\wedge\omega.
\end{equation}
Then the product manifold $S^1\times V$ has a natural $G_2$ structure whose
$G_2$-form is
\begin{equation} \label{eq:product G2 form}
\Phi:=d\varphi \wedge \omega + \operatorname{Re}(\Omega),
\end{equation}
where $\varphi$ is an angular coordinate on the circle.\footnote{Notice that
the phase of $\Omega$ can be varied, which varies the $G_2$ structure
on $S^1\times V$.}

To see this, we let $z_1$, $z_2$, $z_3$ be complex coordinates on $V$
for which $\Omega=dz_1\wedge dz_2\wedge dz_3$ and
$\omega=\frac i2\left(dz_1\wedge \overline{dz_1} + dz_2\wedge \overline{dz_2} + dz_3\wedge \overline{dz_3} \right)$.  We let $\varphi=x_1$, $z_1=x_2+ix_3$,
$z_2=x_4+ix_5$, $z_3=x_6+ix_7$.  Then a brief calculation gives
\begin{equation}
\begin{gathered}
 \Omega = (dx_2+idx_3)\wedge (dx_4+idx_5) \wedge (dx_6+idx_7)\\
 \operatorname{Re}(\Omega) = dx_{246} - dx_{257} - dx_{347} - dx_{356}\\
 \omega = dx_2\wedge dx_3 + dx_4\wedge dx_5 + dx_6\wedge dx_7\\
 d\varphi \wedge \omega = dx_{123} + dx_{145} + dx_{167}.\\
\end{gathered}
\end{equation}
It follows that $d\varphi \wedge \omega + \operatorname{Re}(\Omega)$ is
a $G_2$-form.  Of course, the holonomy on the product manifold $S^1\times V$
is actually a subgroup of $SU(3)$ rather than being all of $G_2$.

A variant of this construction leads to the ``barely $G_2$ manifolds''
studied by Joyce \cite{MR1424428} and Harvey--Moore \cite{Harvey:1999as}:  if $V$ has an
anti-holomorphic involution $\alpha$ which maps $\omega\mapsto -\omega$
and $\Omega\to \overline{\Omega}$, then $(-1,\alpha)$ preserves the
$G_2$ form on $S^1\times V$ and so leads to a $G_2$ structure on
the quotient $(S^1\times V)/(-1,\alpha)$.  If $\alpha$ has no fixed points,
then this quotient is again a seven-manifold.  (This is a case with
holonomy contained in $SU(3)\rtimes\mathbb{Z}_2$ rather than being all of $G_2$.)

Finally, one of the key building blocks for the TCS construction is
a $G_2$ manifold $S^1\times V$ in which $V$ is itself  the product of
$\mathbb{C}^*$ with
a K3 surface $S$.
To define the Calabi--Yau structure on $V$,
we must specify both a Ricci-flat K\"ahler form
$\omega_S$ and a holomorphic $2$-form $\Omega_S$ on the K3 surface $S$,
and for this purpose we use the
normalization $\Omega_S\wedge \overline{\Omega_S} = 2\omega_S\wedge \omega_S$
which implies
\begin{equation}
\operatorname{Re}(\Omega_S)\wedge \operatorname{Re}(\Omega_S) =
\operatorname{Im}(\Omega_S)\wedge \operatorname{Im}(\Omega_S)=
\omega_S\wedge \omega_S.
\end{equation}
This is the normalization familiar in hyperK\"ahler geometry, because in
this case the triple
$(\omega_S,\operatorname{Re}(\Omega_S),\operatorname{Im}(\Omega_S))$
is an orthogonal basis of the space of  self-dual harmonic $2$-forms on $S$,
and all basis elements have the
same norm in $H^2(S,\mathbb{R})$ (in a suitable normalization).  
In fact, given any rotation
in $SO(3)$, we can change
the complex structure on $S$ without changing the underlying Ricci-flat
metric in such a way as to apply the given rotation to the basis
$(\omega_S,\operatorname{Re}(\Omega_S),\operatorname{Im}(\Omega_S))$.

We now choose a complex linear coordinate $z=e^{t+i\theta}$
on $\mathbb{C}^*$
and define, on $V=\mathbb{C}^*\times S$,
\begin{equation}
\label{eq:CY cyl structure}
\begin{aligned}
\omega &= \frac{i\,dz\wedge \overline{dz}}{2z\overline{z}} + \omega_S
=  dt \wedge d\theta +\omega_S \\
\Omega &= -i\, \frac{dz}z \wedge \Omega_S  = (d\theta - i\,dt)
\wedge\Omega_S,\\
\end{aligned}
\end{equation}
so that
\begin{equation}
\operatorname{Re}(\Omega) = d\theta\wedge \operatorname{Re}(\Omega_S)
+dt \wedge \operatorname{Im}(\Omega_S).
\end{equation}

Such a $V$, equipped  with $\omega$ and $\Omega$, is called a
{\em Calabi--Yau cylinder},
and the map $\xi:V\to\mathbb{R}$
defined by $\xi(z,x)=\log |z|$ is called the {\em cylinder
projection}.\footnote{In earlier papers, the term
``Calabi--Yau cylinder'' was used for only half of this space,
namely, $\xi^{-1}(0,\infty)$.}

For a Calabi--Yau cylinder $V$, the three-form
\begin{equation}
\Phi = d\varphi\wedge dt \wedge d\theta
+d\varphi\wedge \omega_S + d\theta\wedge \operatorname{Re}(\Omega_S)
+dt \wedge \operatorname{Im}(\Omega_S).
\end{equation}
on $S^1\times V$ defines
a $G_2$ structure\footnote{To
verify the normalization condition, we define
$\omega_0 = \frac{i\,dz\wedge \overline{dz}}{2z\overline{z}}$
so that $\omega = \omega_0+\omega_S$ and compute:
$$
\frac i8\Omega \wedge \overline{\Omega} = \frac i8\,\frac{dz}{z} \wedge \frac{\overline{dz}}{\overline{z}}
\wedge \Omega_S \wedge \overline{\Omega_S}
=\frac14 \omega_0\wedge \Omega_S \wedge \overline{\Omega_S}
=\frac12 \omega_0\wedge\omega_S\wedge\omega_S = \frac16\omega\wedge\omega\wedge\omega.
$$.}
with
a very interesting property which is the basis
of the TCS construction.  Because the Ricci-flat metric on $S$ is
hyperK\"ahler,
we can change the complex structure on $S$ (without changing the underlying
Ricci flat metric) to obtain a new K3 surface $\Sigma$ with K\"ahler form
$\omega_\Sigma$ and holomorphic $2$-form $\Omega_\Sigma$ such that
$\omega_\Sigma=\operatorname{Re}(\Omega_S)$, $\operatorname{Re}(\Omega_\Sigma)=\omega_S$, and $\operatorname{Im}(\Omega_\Sigma) = - \operatorname{Im}(\Omega_S)$.
Then if we send $(\varphi,t,\theta,S)$ to $(\theta,-t,\varphi,\Sigma)$, the $G_2$ structure is unchanged!

\subsection*{Preliminaries for Twisted Connected Sums}

We will need the notion of an \emph{asymptotically cylindrical} (ACyl)
Calabi-Yau threefold, but before giving the detailed definitions we would
like to state the basic idea. The TCS construction of compact $G_2$
manifolds utilizes two complex threefolds which ``asymptote'' in a
particular way that allows for a particular gluing procedure. The
complex threefold will be an ACyl Calabi-Yau threefold, which is
defined to asymptote to a Calabi-Yau cylinder.

Let
$V$
be a complete, but not necessarily compact,
Calabi-Yau threefold on which a Ricci-flat K\"ahler form $\omega$ and a
holomorphic three-form $\Omega$ have been specified. We say $V$ is an
\emph{asymptotically cylindrical (ACyl) Calabi-Yau threefold} if there is a
compact set $K\subset V$, a Calabi-Yau cylinder $V_\infty$ with
cylinder projection $\xi_\infty:V_\infty\to\mathbb{R}$, and a
diffeomorphism $\eta:\xi_\infty^{-1}(0,\infty)\rightarrow V\setminus K$ such that
$\forall k \ge 0$, some $\lambda > 0$, and as $t\rightarrow \infty$
\begin{align}
  \eta^*\omega-\omega_\infty = d\rho, \qquad \text{for some } \rho \text{ such that } |\nabla^k \rho| = O(e^{-\lambda t}) \nonumber \\
  \eta^*\Omega - \Omega_\infty = d\zeta, \qquad \text{for some } \zeta \text{ such that } |\nabla^k \zeta| = O(e^{-\lambda t})
\end{align}
where $\nabla$ and $|\cdot|$ are defined using the Calabi--Yau metric
$g_\infty$ on $V_\infty$. We refer to $V_\infty=\mathbb{R}^+ \times S^1 \times S$ as
the \emph{asymptotic end} of $V$ and the associated hyperK\" ahler K3 surface $S$ as the \emph{asymptotic K3 surface} of V.

\sskip Since the TCS is a powerful construction technique for building compact
$G_2$ manifolds from elementary parts, it is of fundamental importance
to be able to construct the parts themselves. Namely, we would like to
have a theorem specifying how to construct ACyl Calabi-Yau threefolds
from K\" ahler threefolds in a simple way. This is as follows \cite{MR3109862}. Let $Z$
be a closed K\" ahler threefold with a morphism $f:Z\rightarrow \bP^1$
that has a reduced smooth K3 fiber $S$ with class $[S]=-[K_Z]$, and
let $V=Z\setminus S$. If $\Omega_S$ is a nowhere vanishing
$(2,0)$-form on $S$ and the K\" ahler form $\omega_S$ 
is the restriction of a K\" ahler class on $Z$, then $V$
has a metric that makes it into an ACyl Calabi-Yau threefold which asymptotes to a Calabi-Yau
cylinder satisfying (\ref{eq:CY cyl structure}).

In addition to this, Corti Haskins Nordstr\" om and Pacini
\cite{Corti:2012kd} make some additional assumptions that simplify the
calculation of topological invariants for their $G_2$ manifolds. To
this end, let $Z$ be a nonsingular algebraic 3-fold $Z$ together with
a projective morphism $f:Z\rightarrow \bP^1$. 
Such a $Z$ is a \emph{building
  block} if:

\vspace{.25cm}
(i) the anticanonical class $-K_Z\in H^2(Z)$ is primitive.

(ii) $S=f^*(\infty)$ is a nonsingular K3 surface in the anticanonical class.

(iii) The cokernel of the restriction map $H^2(Z,\mathbb{Z})\to H^2(S,\mathbb{Z})$ is torsion-free.

(iv) The group $H^3(Z)$, and thus also $H^4(Z)$, is torsion-free.

\vspace{.25cm}
\noindent The original building blocks of Kovalev \cite{KovalevTCS} were Fano
threefolds, while Kovalev-Lee \cite{KovalevLee} utilized building
blocks with non-symplectic involutions on the K3s. The broadest class
of building blocks to date utilize weak-Fano three-folds due to CHNP
\cite{MR3109862,Corti:2012kd}. These will be reviewed momentarily, but
let us first introduce the TCS construction since it does not require
a specific type of building block.

\subsection*{Compact $G_2$ Manifolds from Twisted Connected Sums}
\label{sec:G2 TCS}
We now review Kovalev's twisted connected sum construction for compact
$G_2$ manifolds. The basic idea is to glue two ACyl Calabi-Yau
threefolds in a particular way which ensures the existence of a $G_2$
metric.  To do so, the ACyl Calabi-Yau threefolds must be compatible.

Let $V_\pm$ be a pair of asymptotically cylindrical Calabi-Yau
threefolds with K\"ahler forms $\omega_\pm$ and holomorphic three-forms
$\Omega_\pm$ specified.
Then by definition $V_\pm$ asymptotes to one end of a Calabi-Yau cylinder,
i.e., $V_{\infty,\pm} = \bR^+ \times S^1_\pm \times S_\pm$ where
$S_\pm$ is the asymptotic hyperK\" ahler $K3$ surface of $V_\pm$.  Of
course, these are real six-manifolds, and we must add a seventh
dimension and glue appropriately.  To add the seventh dimension,
define the seven-manifolds $M_\pm = S^1_\mp \times V_\pm$ and let
$\theta_\mp$ be the standard coordinate on the $S^1$. Since $V_\pm$
asymptotes to a Calabi-Yau cylinder, $M_\pm$ asymptotes to a circle
product with a Calabi-Yau cylinder. Now suppose that there exists a
diffeomorphism $r:S_+\to S_-$, preserving the Ricci-flat metric,
such that
\begin{equation} \label{eq: Donaldson matching}
\begin{aligned}
r^*(\omega_{S_-}) &= \operatorname{Re}(\Omega_{S_+}) \\
r^*(\operatorname{Re}(\Omega_{S_-})) &= \omega_{S_+} \\
r^*(\operatorname{Im}(\Omega_{S_-})) &=-\operatorname{Im}(\Omega_{S_+}) \\
\end{aligned}
\end{equation}
Then we can glue the seven-manifolds $M_\pm$ in
their asymptotic regions as follows: on the region in $\bR^+$ defined
by $t\in (T,T+1)$ consider the diffeomorphism
\begin{align}
F: \qquad M_+ \cong S^1_- \times \bR^+ \times S^1_+ \times S_+ \qquad &\longrightarrow \qquad
S^1_+ \times \bR^+ \times S^1_- \times S_- \cong M_-, \nonumber \\
(\theta_-,t,\theta_+,x) \qquad &\longmapsto  \qquad (\theta_+,T+1-t,\theta_-,r(x))
\end{align}
There are $G_2$ structures on these asymptotic regions; see
e.g. \cite{Corti:2012kd} for their detailed structure, since they
will not be critical for us. By truncating each $M_\pm$ at $t=T+1$ we
obtain a pair of compact seven-manifolds $M_\pm(T)$ with boundaries
$S^1_+\times S^1_- \times S_\pm$; then they can be glued together with the
diffeomorphism $F$ to form a \emph{twisted connected sum}
seven-manifold $M_r = M_+(T) \cup_F M_-(T)$. This is a compact
seven-manifold which admits a closed $G_2$ structure that is
determined by the $G_2$ structures on $M_\pm$; however, they
are not a priori torsion-free. This leads to:

\vspace{.25cm}
\textbf{\emph{Kovalev's Theorem:}} Let $(V_\pm,\omega_\pm,\Omega_\pm)$ be
two ACyl Calabi-Yau three-folds with asymptotic

ends of the form
$\bR^+\times S^1 \times S_\pm$ for a pair of hyperK\" ahler K3
surfaces $S_\pm$, and suppose that

there exists a diffeomorphism $r:S_+ \rightarrow
S_-$ preserving the Ricci-flat metrics and satisfying

\ref{eq: Donaldson matching}. Define the twisted connected sum $M_r$ as
above with closed $G_2$ structure $\Phi_{T,r}$. Then for

sufficiently
large $T$ there is a torsion-free
perturbation of $\Phi_{T,r}$ within its cohomology class;

call this
torsion-free $G_2$ structure $\Phi$.

\vspace{.25cm}
\noindent Since a torsion-free $G_2$ structure determines a metric
with holonomy exactly $G_2$, $(M_r,\Phi)$ is a compact seven-manifold
with holonomy $G_2$.

We call a diffeomorphism $r:S_+ \rightarrow
S_-$ preserving the Ricci-flat metrics and satisfying
\ref{eq: Donaldson matching} a {\em \KM}.\footnote{This is usually called
a ``hyper-K\"ahler rotation'' in the literature,
but in fact it is a very particular
type of hyper-K\"ahler rotation and we prefer a different name.  
According to \S11.9 of \cite{joyce2000compact}, the use of such a diffeomorphism
for a gluing construction of $G_2$ manifolds
  was first proposed by Donaldson. }

\sskip In summary, to build twisted connected sum $G_2$ manifolds, one
needs only appropriate building blocks $Z_\pm$ and to choose an
appropriate \KM\  on associated asymptotic
hyperK\" ahler K3 surfaces $S_\pm$. One aspect of the recent
progress\cite{Corti:2012kd} by Corti, Haskins, Nordstr\" om, and
Pacini was to greatly enlarge the known set of building blocks
compared to those originally considered by Kovalev \cite{KovalevTCS},
who utilized \emph{building blocks of Fano type} built from Fano
threefolds, and by Kovalev-Lee \cite{KovalevLee}, who utilized
\emph{building blocks of non-symplectic type}; the new larger class of
building blocks of \cite{Corti:2012kd} utilize weak-Fano threefolds.

Let us explain how the authors of \cite{Corti:2012kd} obtain ACyl Calabi-Yau threefolds from
semi-Fano threefolds. First, a \emph{weak Fano threefold} is a
nonsingular projective complex threefold $Y$ such that the
anticanonical class $-K_Y$ satisfies $-K_Y\cdot C \ge 0$ for any
compact algebraic curve $C\subset Y$, and furthermore $(-K_Y)^3 >
0$. Since the latter is an even integer, the \emph{anticanonical
  degree} $(-K_Y)^3$ can be defined in terms of the genus $g_Y$ of $Y$
as $(-K_Y)^3=:2g_Y - 2$. There is a key fact about any smooth weak
Fano threefold that is important for constructing building blocks: a
general divisor in the anticanonical class is a non-singular K3
surface. CHNP make the additional assumption that the linear system
$|-K_Y|$ contains two non-singular members $S_0$ and $S_\infty$
which intersect transversally. Most weak Fano threefolds satisfy
this assumption.

A \emph{building block of semi-Fano type} is defined as followed. Let $Y$ be a
semi-Fano threefold with torsion-free $H^3(Y)$, $|S_0,S_\infty|\subset
|-K_Y|$ a generic pencil with smooth base locus $C$, and take $S \in
|S_0,S_\infty|$ generic. Furthermore, let $Z$ be the blow-up of $Y$ at
$C$. Then $S$ is a smooth K3 surface and its proper transform in $Z$
is isomorphic to $S$. The pair $(Z,S)$ constructed in this way
is called a semi-Fano building block. Then the image 
of
$H^2(Z,\mathbb{Z}) \rightarrow H^2(S,\mathbb{Z})$ equals that of 
$H^2(Y,\mathbb{Z})\rightarrow H^2(S,\mathbb{Z})$,
and furthermore the latter map is injective; this will be important
in a theorem on the cohomology of the $G_2$ manifold which we will review
in the next section. Other relevant technical statements and remarks
can be found in section 3 of \cite{Corti:2012kd}.

\sskip

\subsection*{Topology of the Twisted Connected Sum $G_2$ manifolds }

Studying the topology of a TCS $G_2$ manifold $X$ is critical for
understanding the physics of the associated M-theory vacuum.  As
$X$ is constructed from elementary building blocks, its topology is
determined by the topology of the building blocks and the gluing
map. Though most of the discussion holds for general building blocks,
we will occasionally comment on results specific to the use of
building blocks of semi-Fano type. For more details
see section $4$ of \cite{Corti:2012kd}.

The fundamental group and the Betti numbers were computed for early
examples of $G_2$ manifolds, but thanks to \cite{Corti:2012kd},
it is now possible to compute the
full integral cohomology for many twisted connected sums, including
the torsional components of $H^3(X,\bZ)$ and $H^4(X,\bZ)$, as well as
the first Pontryagin class $p_1$. If not for a general
observation that we will discuss, the explicit knowledge of $p_1$ in
examples would play a critical role \cite{Witten:1996md} in determining
the quantization of M-theory flux in those examples.

The second integral cohomology of a twisted connected sum $G_2$
manifold $X$ is given by
\begin{equation}
H^2(X,\bZ) = (N_+ \cap N_-)\oplus K_+ \oplus K_-
\label{eqn:H2}
\end{equation}
where $N_\pm$ is the image of $H^2(Z_\pm,\bZ)$ in $H^2(S_\pm,\bZ)$, and $K_\pm:=ker(\rho_\pm)$
where we have
\begin{equation}
\rho_\pm :\qquad  H^2(V_\pm,\bZ)\longrightarrow H^2(S_\pm,\bZ)
\end{equation}
being the natural restriction maps.\footnote{There is a neighborhood of
$S$ in $Z$ which is diffeomorphic to a product of $S$ with a disk.   The
cohomology groups of all the nearby K3 surfaces to $S$ in this
neighborhood can be identified with those of $S$, so restricting
a cohomology class to any one of those nearby K3 surfaces gives
a cohomology class on $S$ itself.}
Intuitively, the
contributions of $K_\pm$ to $H^2(X,\bZ)$ are non-trivial classes on
the ACyl Calabi-Yau threefolds $V_\pm$ which restrict trivially to the
K3 surfaces $S_\pm$, and therefore the gluing map (which twists the
classes of the K3 surfaces according to the \KM\ $r$)
will not affect elements of $K_\pm$; they become non-trivial in $X$, as well.
Alternatively, classes which restrict non-trivially to the K3's are subject to
the gluing map, and therefore only classes in the intersection $N_+ \cap N_-$,
become non-trivial classes in $H^2(X,\bZ)$.

For our purposes we will not need to know the full third cohomology of $X$, instead
only that it contains three-forms related to the building blocks
\begin{equation}
H^3(X,\bZ) \supset H^3(Z_+,\bZ) \oplus H^3(Z_-,\bZ)\oplus K_+ \oplus K-
\label{eqn:H3}
\end{equation}
and we refer the reader to Theorem $4.9$ of \cite{Corti:2012kd} for the full result. This
is an interesting result: for any $\alpha_\pm \in K_\pm$ on one of the building blocks,
we have an associated non-trivial two-form and threeform on $X$, arising as \cite{Corti:2012kd}
\begin{equation}
\alpha_\pm \in K_\pm \qquad \qquad \longleftrightarrow \qquad \qquad \alpha_\pm \in H^2(X,\bZ) \qquad \text{and} \qquad \alpha_\pm\wedge d\theta_\mp \in H^3(X,\bZ)
\end{equation}
where again $\theta_\mp$ is the coordinate on $S^1_\mp$. So it is precisely clear how a non-trivial
two-form on a building block can give both non-trivial two-forms and three-forms on $X$.

Finally, we will utilize some results of \cite{Corti:2012kd} regarding
the existence of associative submanifolds, both rigid and not.  There
are two known ways in which to obtain associative submanifolds in TCS
$G_2$ manifolds.  Recall that $V$ is an ACyl Calabi-Yau threefold $V$
of a TCS $G_2$ building block. The first result is that if $L\subset V$
is a compact special Lagrangian submanifold with $b_1(L)=0$ and $L$ is
non-trivial in the relative homology $H_3(V,S^1 \times S)$, then there
is a small deformation of $L$ in $X$ which is an associative
threefold. This associative is not rigid.  The second result is that
if $C$ is a rigid holomorphic curve in $V$, then a small deformation
of $S^1\times C$ in $X$ is a rigid associative. This latter result is
significant. It gives the first construction technique for compact
rigid associative submanifolds in compact $G_2$ manifolds, and
therefore it is now possible to compute the form of membrane instanton corrections
to the superpotential in examples; see the following.

\section{A Rich Example}
\label{sec:example}

In this section we study an explicit example of
\cite{Corti:2012kd}, performing a number of new computations necessary
to uncover interesting physical aspects of this M-theory vacuum, as well
as studying topology changing transitions to other $G_2$ manifolds and M-theory vacua.

For the $G_2$ manifold $X$ that we study, we will show that M-theory
on $X$ yields an $\cN=1$ supersymmetric four-dimensional supergravity
theory at low energies with $U(1)^3$ gauge symmetry and a spectrum of
massive charged particles including trifundamentals. Vacua exhibiting
$U(1)^3$ gauge symmetry and trifundamental matter were also recently
discovered \cite{Cvetic:2013qsa} among F-theory
compactifications. Using the topological progress of
\cite{Corti:2012kd}, we also compute the form of membrane
instanton corrections to the superpotential (for the first time in a
compact model) and $G_2$ topology changing transitions. The
transitions of $X$ to other $G_2$ manifolds include both the
(non-isolated) $G_2$ flop and $G_2$ conifold transitions; in the
former \emph{both} two-cycles and three-cycles collapse and re-emerge
in a different topology, while in the latter two-cycles and
three-cycles collapse, but only a three-cycle emerges after
deformation. Physically, the non-isolated conifold transition breaks
the $U(1)^3$ gauge symmetry of M-theory on $X$ to $U(1)^2$ in the
usual way.

\sskip \sskip Before delving into details, we would like to state the
basic mathematical idea that gives rise to the interesting physics. We will study an
example from \cite{Corti:2012kd} which utilizes one building block
with $K\ne 0$, which is known to fit into a matching pair giving rise
to a $G_2$ manifold with $H^2(X,\bZ)=K\cong \bZ^3$, and therefore the
associated M-theory vacuum exhibits $U(1)^3$ gauge symmetry. We will
compute the topological intersections of two-cycles with five-cycles
in the $G_2$ manifold to determine the charges of massive particles on
this M-theory vacuum; in this example these intersections are
conveniently related to intersections in the algebraic threefold of the
building block. Given the homology classes of some rigid holomorphic
curves we will determine the homology classes of their associated
rigid associative threefolds; this determines the moduli dependence of
some instanton corrections to the superpotential, and we find a six
term generalized racetrack in three moduli. We will then study topology
change in detail, where the non-isolated $G_2$ flop and conifold transitions occur
via movement in $G_2$ moduli and can be understood in terms of induced
flop and conifold transitions in the building block.

As a brief physical review, consider a compactification of M-theory on a smooth
$G_2$ manifold $X$ at large volume\footnote{Outside of the strict large volume
approximation, M-theory compactifications with non-abelian gauge sectors are sometimes well approximated by
a combined supergravity and super Yang-Mills action; see \cite{Acharya:2000gb,Friedmann:2002ty}.}. Its metric is determined by a
torsion-free $G_2$ form $\Phi \in H^3(X)$, and $\Psi \equiv \star
\Phi$ is the dual four-form. This compactification gives a
four-dimensional $\cN=1$ theory with an associated massless effective
action obtained from Kaluza-Klein reduction of 11-dimensional
supergravity; see \cite{Papadopoulos:1995da} for more details. It exhibits
\begin{center}
  $b_2(X)$ abelian vector multiplets from $C_3$-reduction along $\sigma \in H^2(X)$\\ \vspace{.1cm} and \vspace{.1cm} \\
  $b_3(X)$ neutral chiral multiplets from $\int_T(\Phi + i C_3)$
  for all $T \in H_3(X)$,
\end{center}
where $C_3$ is the M-theory three-form. We emphasize that the gauge
group is $G=U(1)^{b_2(X)}$; it is an abelian theory without any
massless charged particles; however, massive charged particles can arise
from M2-branes wrapped on two-cycles.

\subsection{The $G_2$ Manifold and Relevant Building Block}
\label{sec:relevant building block}

We wish to study an example from \cite{Corti:2012kd} where $b_2(X)\ne
0$ arises from the fact that one of the building blocks has $K\ne
0$. While we will focus mostly on a particular building block of that
type since it gives rise to the physics we are interested in, it is
worth noting that it does form a matching pair with building blocks
from example $7.1$ of \cite{Corti:2012kd}; the associated $G_2$
manifolds have $H^2(X,\bZ)=K_+$, where $K_+$ is the $K$ lattice of the
building block we study in detail. From now on we will drop $\pm$
subscripts, focusing only on the building block of interest.

As discussed, one way to obtain a building block with $K\ne 0$ is to
blow up an algebraic threefold along a non-generic (rather than
generic) anticanonical pencil.  The example we use is example 4.8 of
\cite{MR3109862}, and the threefold we begin with is the simplest one,
$Y=\bP^3$. Consider the non-generic pencil $|S_0,S_\infty|\subset
|\cO(4)|$ with $S_0$ the tetrahedron $S=\{x_1x_2x_3x_4=0\}$ and
$S_\infty$ a generic non-singular quartic surface which meets all
coordinate planes $x_i=0$ transversely. The base locus of the pencil
is the union of four non-singular curves $C_i:= \{x_i=0\} \cap
S_\infty$ where since
\begin{equation}
\chi(T C_i) =\int_{C_i} c_1(TC_i) = \int_{4H^2}(-H) = -4 =2-2g
\end{equation}
we see that $C_i$ is a genus $3$ curve. $Z$ is obtained from $Y$ by blowing
up the base curves $C_i$ one at a time, and the associated ACyl Calabi-Yau
threefold is $V=Z\setminus S$.

For simplicity let $Z$ be
obtained by blowing up along $C_1,C_2,C_3,C_4$ in that
order. Associated to these blow-ups are four exceptional divisors
$E_i$, giving $h^2(Z) = h^2(Y) + 4 = 5$. After the first blowup, the
geometry appears as
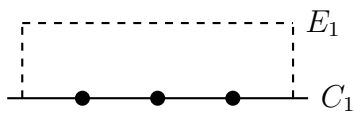
\begin{center}
  \begin{tikzpicture}
    \draw[thick] (-20mm,0cm) -- (20mm,0cm);
    \draw[thick,dashed] (-18mm,0mm) -- (-18mm,10mm);
    \draw[thick,dashed] (18mm,0mm) -- (18mm,10mm);
    \draw[thick,dashed] (-18mm,10mm) -- (18mm,10mm);
    \fill (-10mm,0mm) circle (1mm);
    \fill (0mm,0mm) circle (1mm);
    \fill (10mm,0mm) circle (1mm);
    \node at (24mm,0mm) {$C_1$};
    \node at (22mm,10mm) {$E_1$};
  \end{tikzpicture}
\end{center}
where the dashed box denotes the exceptional divisor $E_1$ obtained by
blowing up up along $C_1$ and the three dots represent the intersections
of $C_1$ with $C_2$, $C_3$, and $C_4$, four times each. Blowing up
again, we obtain
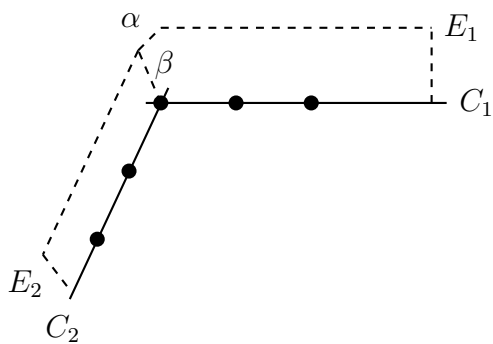
\begin{center}
  \begin{tikzpicture}
    \draw[thick] (-20mm,0cm) -- (20mm,0cm);
    \draw[thick] (-17mm,2mm) -- +(245:31mm);
    \draw[thick,dashed] (-18mm,0mm) -- (-21mm,7mm);
    \draw[thick,dashed] (-21mm,7mm) -- (-18mm,10mm);
    \draw[thick,dashed] (-21mm,7mm) -- +(245:30mm);
    \draw[thick,dashed] (-21mm,7mm)+(245:30mm) -- (-30mm,-25mm);
    \draw[thick,dashed] (18mm,0mm) -- (18mm,10mm);
    \draw[thick,dashed] (-18mm,10mm) -- (18mm,10mm);
    \fill (-18mm,0mm) circle (1mm);
    \fill (-18mm,0mm)+(10mm,0mm) circle (1mm);
    \fill (-18mm,0mm)+(20mm,0mm) circle (1mm);
    \fill (-18mm,0mm)+(245:10mm) circle (1mm);
    \fill (-18mm,0mm)+(245:20mm) circle (1mm);
    \node at (24mm,0mm) {$C_1$};
    \node at (22mm,10mm) {$E_1$};
    \node at (-31mm,-30mm) {$C_2$};
    \node at (-36mm,-24mm) {$E_2$};
    \node at (-22mm,11mm) {$\alpha$};
    \node at (-17.5mm,5mm) {$\beta$};
    \end{tikzpicture}
\end{center}
where now we see $E_1$ and $E_2$, the exceptional divisors of the
consecutive blow-ups along $C_1$ and $C_2$. $E_1$ and $E_2$ are
fibrations over $C_1$ and $C_2$ with generic fibers being curves of
class $\gamma_1$ and $\gamma_2$. The dot at the intersection of $C_1$
and $C_2$ represents their four intersection points, and the
additional dots on $C_1$ and $C_2$ represent their four intersections
with $C_3$ and $C_4$ respectively.  The jagged dashed curve represents
the inverse image of $C_1\cdot C_2$, which is a singular curve that is
a reducible variety with two components $\alpha$ and $\beta$; these
are curves of class $\gamma_1-\gamma_2$ and $\gamma_2$, respectively,
In fact since $C_1\cdot C_2$ occurs at four points there are actually
four rigid holomorphic curves of class $\gamma_1-\gamma_2$. We have
shown these images primarily to demonstrate the appearance of such
holomorphic curves, but also to give intuition for the
geometry. Blow-ups three and four proceed in a similar fashion, but
are harder to draw.

After performing all of the blow-ups, we would like to know the
effective curves in $Z$ and their intersections with divisors. $E_i$
is an exceptional divisor that is fibered over $C_i$ with generic
fiber a $\bP^1$ $\gamma_i$ that moves in families. Above the points
$C_i\cdot C_{j>i}$ the fiber is a rigid holomorphic curve of class
$\gamma_i-\gamma_{j>i}$; given the six possible choices of $i,j$ and
the fact that $C_i\cdot C_j$ is a set of four points, this yields
curves in six homology classes with four representatives each, for a
total of $24$ rigid holomorphic curves. Note that none of these curve
classes can be written as a positive linear combination of two others.

How do the exceptional divisors $E_i$ intersect the curves $\gamma_j$?
Choose a general fiber in $E_j$; this is a curve of class $\gamma_j$
and it clearly does not intersect any exceptional divisor $E_{i\ne
  j}$.  On the other hand, since the rigid curves of class
$\gamma_2-\gamma_1$ are contained in $E_1$ and are transverse to
$E_2$, so $E_2\cdot(\gamma_1 - \gamma_2)=1$ and therefore $E_2\cdot
\gamma_2=-1$. However, computing $E_1\cdot \gamma_1$ cannot be done by
counting points and must be done indirectly. To do so we use a few
simple facts. First, $\gamma_1$ is a rational curve contained in
$E_1$, and therefore $-\chi(\gamma_1) = 2g-2=-2$. Alternatively
$-\chi(\gamma_1) = -\int_{\gamma_1} c_1(\gamma_1) =
(K_{E_1}+\gamma_1)\cdot\gamma_1=K_{E_1}\cdot \gamma_1$ where the last
equality holds because $\gamma_1$ moves in $E_1$. Letting our blow-up
be $\pi:Z\to Y$, then $K_Z = \pi^*K_Y + E_1 + E_2+ E_3+
E_4$ and adjunction therefore gives $K_{E_1} = (K_Z + E_1)|_{E_1} = (\pi^*(K_Y) + 2E_1 + E_2 + E_3 + E_4)|_{E_1}$.
Putting it all together
\begin{equation}
  -2 = -\chi(\gamma_1) = (\pi^*(K_Y) + 2E_1 + E_2 + E_3 + E_4)|_{E_1} \cdot \gamma_1 = 2E_1\, \cdot \gamma_1
\end{equation}
The last equality holds because $E_{j>1}\cdot \gamma_1=0$ from above
and also since a generic canonical divisor in $Y$ misses a generic
point in $C_1$ and therefore the rational curve of class $\gamma_1$
above such a point in the blowup $Z$. We therefore obtain
$E_1\cdot \gamma_1 = -1$ and overall have
\begin{equation}
E_i\cdot \gamma_j = -\delta_{ij}
\label{eq:intersections}
\end{equation}
which we will use to compute physically relevant intersections in a
moment.

Before doing so, we compute $K$ in order to determine the number of $U(1)$ symmetries
and their generators (in cohomology).  Recalling that $V=Z\setminus S$ and $K=ker(\rho)$ with
\begin{equation}
\rho:\,H^2(V)\rightarrow H^2(S),
\end{equation}
the restriction map and where $b_2(V)=4$ since we have subtracted out
$S$. Now, since each $E_i$ is a fibration over a curve $C_i$ which
itself is a curve in $S$ of class $H|_S$, then $\rho(E_i)=H|_S$ and
therefore $\rho(E_i-E_j)=0\in H^2(S)$; i.e., $E_i-E_j\in K$ and in fact
we will choose a basis $E_1-E_2$, $E_1-E_3$ and $E_1-E_4$ for $K$;
call these $D_1$, $D_2$, and $D_3$ respectively. We also see that $K$
is rank three, and since the second cohomology of a TCS $G_2$ manifold
$H^2(X,\bZ) \supset K$ then we have at least three $U(1)$ symmetries
(and in fact choosing the other building block as in
\cite{Corti:2012kd} we will have precisely three.)  By choosing
generators of $K$ we have chosen a basis for the three associated
$U(1)$'s in the M-theory compactification. One way to see this is to
note that $D_i\times S^1$ now are three non-trivial five-cycles in $X$
which have dual non-trivial two forms, which give rise to $U(1)$ symmetries.
We  would like to compute the
intersection of five-cycles with two-cycles since these
determine the charges of massive particles in $G_2$ compactifications
if positive volume cycles exist. For example, for us the positive
two-cycles are the ones obtained from holomorphic curves in the
building block.

Since the two-cycles and associated five-cycles in $X$ we are studying
come ``from one end'' of the $G_2$ manifold, i.e., from one building
block rather than from the intersection $N_+\cap N_-$ of the $N$
lattices of the two different building blocks, we can compute the
intersections of these five-cycles and two-cycles one one
end. These intersections in $X$ are determined by
intersections of the relevant divisors and curves in
$V$. Additionally, since the divisors we are interested in generate
$K$ they do not intersect $S$, and therefore any intersection with a
curve $\gamma$ happens away from $S$, so that in all
\begin{equation}
(D_i\times S^1)\cdot_X \gamma = D_i\cdot_V \gamma = D_i\cdot_Z \gamma
\end{equation}
and thus we simply need to compute intersections in $Z$.

\subsection{Massive Charged Particles and Instanton Corrections}
\label{sec:massive charged particles on X}

Let us now study the charged particles in the theory.  These arise
from M2-branes wrapped two-cycles in $X$; since $E_i$ contains
$\gamma_i$ and only differences of the $E_i$'s are in $K$, curves in
$Z$ of class $\gamma_i$ do not become two-cycles in $X$. The rigid
holomorphic curves of class $\gamma_i-\gamma_{j>1}$ do become
two-cycles in $X$, however.  An M2-brane on a curve $\gamma$ gives a
particle of charge $(D_i\times S^1)\cdot_X \gamma=D_i\cdot_Z \gamma=:
Q_i$ under $U(1)_i$. Using (\ref{eq:intersections}) and naming the
particles arising from an M2-brane on a two-cycle of class $\gamma_i-\gamma_j$
to be $\Psi^k_{ij}$ with $k=1,\dots,4$, we compute the charges
\begin{center}
  \begin{tabular}{c|c|c|c|}
    & $Q_1$ & $Q_2$ & $Q_3$ \\ \hline
 $\Psi^k_{12}$ & $-2$ & $-1$  & $-1$ \\ \hline
 $\Psi^k_{13}$ & $-1$ & $-2$ & $-1$ \\ \hline
 $\Psi^k_{14}$ & $-1$ & $-1$ & $-2$ \\ \hline
 $\Psi^k_{23}$ & $1$ & $-1$ & $0$ \\ \hline
 $\Psi^k_{24}$ & $1$ & $0$ & $-1$ \\ \hline
 $\Psi^k_{34}$ & $0$ & $1$ & $-1$ \\ \hline
  \end{tabular}
\end{center}
for these massive particles. As the particles are
massive, they necessarily arise as vector pairs so that for any $\Psi$
there is another chiral multiplet $\bar \Psi$ with opposite charge, so
that the superpotential $W$ contains a term $ m_\Psi \Psi \bar\Psi$. These latter fields arise from
anti M2-branes. Since these particles arise from rigid holomorphic
curves in one of the building blocks, there are representatives of this two-cycle class
within a compact rigid associative submanifold.

There are $24$ rigid holomorphic curves in $V=Z\setminus S$ in six
different homology classes $\gamma_i-\gamma_{j>i}$, with four
representatives of each class. As discussed in section \ref{sec:TCS
  review}, to each such curve there is a compact rigid associative in
$X$, giving $24$ compact rigid associatives in $X$, also in six
different homology classes $T_i-T_{j>i} \in H_3(X,\bZ)$ since the
curves come in six different classes.  M2-branes wrapped on rigid
associative cycles are expected to generate instanton corrections \cite{Harvey:1999as}
to the superpotential\footnote{Determining
    whether or not the instanton corrects the superpotential requires
    a careful analysis of instanton zero modes. The rigidity condition
    ensures the absence of deformation modes that would otherwise kill
    the superpotential correction, but it may be the case in some
    models that the Wilson line modulini associated to the $S^1$ in
    the rigid associative also kill the superpotential
    correction. Whether or not this is the case (in particular whether
    the modulini are lifted by interactions) seems to be model
    dependent, and we leave this analysis to future work.}; for these rigid associatives we have
constructed the associated superpotential takes the
form
\begin{equation}
W \supset 4(A_{1}e^{-\Phi_1}+A_{2}e^{-\Phi_2}+A_{3}e^{-\Phi_3}+A_{4}e^{\Phi_1-\Phi_2}+A_{5}e^{\Phi_1-\Phi_3}+A_{6}e^{\Phi_2-\Phi_3})
\label{eq:explicit instanton corrections}
\end{equation}
where the factor of $4$ is because there are four rigid associatives
(and therefore four instanton corrections) per class and $\Phi_1$,
$\Phi_2$, and $\Phi_3$ are the moduli associated to $T_1-T_2$,
$T_1-T_3$, and $T_1-T_4$, respectively. While there may be other rigid
associatives which also give rise to instanton corrections, we see at
the very least that any TCS $G_2$ manifold constructed from this
building block realizes a six term generalized racetrack in four
different moduli fields.

\subsection{Massless Limits, Topology Change, and the Higgs Mechanism}
\label{sec:massless limits, topology change, and the Higgs mechanism}

In this section we will study singular limits in $G_2$ moduli space
and topology changing transitions. In the singular limit the massive
charged particles of the last section will become massless. In one
transition we will perform a non-isolated $G_2$ flop to another branch of the
moduli space in which these particles are massive; in another we will
perform a non-isolated $G_2$ conifold transition in which one of the $U(1)$
symmetries is broken and there are particles charged under the remaining
$U(1)$ symmetries. More complicated conifolds also exist in this
example. (Non-compact realizations of {\em isolated}\/ $G_2$ flop and conifold
transitions were studied in \cite{Acharya:2000gb,Atiyah:2000zz} and \cite{Atiyah:2001qf},
respectively.)

We discuss a potential technical obstruction before turning to
details. Taking a singular limit in which particles become massless
requires gaining some control over two-cycles: given the lack of a
calibration form for two-cycles, how might one do this? Our basic idea
begins with noting that the rigid associative threefolds in $X$
appeared because of the existence of rigid holomorphic curves in the
building block, and in this example these curves also became
non-trivial two-cycles in $X$. If one flopped a curve in the threefold
building block then sometimes the topology of $X$
itself changes.  Since these curves sit inside rigid associative
threefolds, it is natural to expect that by sending the associative to
zero volume by tuning in $G_2$ moduli, the curve within it might also
collapse. However, since the rigid associatives
of \cite{Corti:2012kd} are diffeomorphic to $S^2\times S^1$, and 
{\em a priori}\/ the
$S^1$ rather than the $S^2$ might collapse. So one would like evidence
in moduli that when the rigid associative threefold collapses, the
$S^2$ within it also collapses. We will argue that this should be
expected when the the (non-isolated) $G_2$ flop and conifold
transitions arise from transitions on one of the building blocks.

\sskip
\sskip
Before studying degenerations of the $G_2$ manifold, we would
like to understand degenerations and topology change in the building block. First note
that the four successive blow-ups of the last section together gave
a birational map $Z\to Y$ that was not crepant; i.e., the canonical class of the variety
changed in the process and therefore $Y$ should be viewed as an auxiliary
variety useful for constructing $Z$, but not related to $V$ in moduli
in the way we would like.

Instead, the variety related to $Z$ which we would like to consider is
$P$, the variety obtained via blowing down all of $24$ rigid
holomorphic curves, which therefore has $24$ conifold points. $Z$ can
be obtained via a sequence of blow-ups along divisors $X_i\equiv \{x_i=0\}$
\begin{equation}
  \label{eq:Pbl}
  Z=P_{4321}\xrightarrow{\pi_1}P_{432}\xrightarrow{\pi_2}P_{43}\xrightarrow{\pi_3}P_4\xrightarrow{\pi_4}P
\end{equation}
where $\pi_i$ is the blow-up along $X_i$. $P$ is simply the total space of the pencil $|S_0,S_\infty|$
\begin{equation}
  \label{eq:P}
  P = \{(x,\lambda) \, |\,  x \in S_\lambda, \text{ and } S_\lambda\in|S_0,S_\infty|\} \subset \bP^3\times \bP^1
\end{equation}
with parameter $\lambda$.  Note here that
each successive blowup adds holomorphic curves in reverse order: in
$Z\to Y$ the successive blow-ups yielded $0$, $4$, $8$, and $12$
holomorphic curves respectively, whereas in $Z\rightarrow P$ the
successive blow-ups yield $12$, $8$, $4$ and $0$ holomorphic curves,
respectively. This is a simple consequence of blowing up along
divisors $X_i$ rather than curves $C_i$. For example, $\pi_4$ blows up
along $X_4$ which contains $12$ conifold points, coming in three sets
of four where $x_4=x_j=0$ for $j=1,2,3$.  Therefore $\pi_4$ produces
$12$ curves, and the next blowup $\pi_3$ along $X_3$ resolves the $8$
conifold points at $x_3=x_k=0$ for $k=1,2$, producing $8$ curves, etc.
The last blow-up giving rise to curves is $\pi_2$ which produces the
curves in class $\gamma_1 - \gamma_2$ in $Z$.

To study topology change, we first blow down to the singular variety
\begin{equation}
Z \xrightarrow{\pi} P_{43}
\end{equation}
where $\pi = \pi_1 \circ \pi_2$. This map blows down the four rigid holomorphic curves in class
$\gamma_1-\gamma_2$ to four conifold points, which are the only
singularities in $P_{43}$. From $P_{43}$, we may perform another small
resolution of the conifold points (i.e., not $\pi$) which flops the
curve $\gamma_1-\gamma_2$, or we may deform the conifold points.

The other small resolution proceeds in the usual way. The divisor we
blow up along, $X_2$, is a non-Cartier Weil divisor that passes
through the conifold points. One such point is locally of the form
\begin{equation}
y x_2  = z w
\end{equation}
and there are two resolutions of the conifold corresponding to blowing up along $x_2=z=0$
or $x_2=w=0$; though codimension two in the local $\bC^4$ ambient space, this blow-up
is codimension one in the hypersurface, i.e., along the divisor $X_2$.

The deformation is more subtle, but can be understood by first thinking of a deformation of $P$ and
relating it to $P_{43}$. Recalling that $P$ is of the form $(x,\lambda)$, for $\lambda =0$ this is $(S_0,\lambda)$
where $S_0=\{x_1x_2x_3x_4\}=0$, we can perform a deformation $P\xrightarrow{\pi_\epsilon} P_\epsilon$ by deforming
$S_0\to S_{0,\epsilon}$ where
\begin{equation}
  \label{eq:Sep}
  S_{0,\epsilon} = \{(x_1x_2+\epsilon \, Q_2) x_3x_4=0\}
\end{equation}
in terms of a quadric $Q_2$ in $x_1,x_2,x_3,x_4$. For simplicity also define $Q\equiv x_1x_2 + \epsilon\, Q_2$
and note that we recover $S_0$ in the $\epsilon \to 0$ limit. Then $P_\epsilon$ is just the total
space of the pencil $|S_{0,\epsilon},S_{\infty}|$
\begin{equation}
  \label{eq:Peps}
  P_\epsilon = \{(x,\lambda) \, |\,  x \in |S_{0,\epsilon},S_\infty|\} \subset \bP^3\times \bP^1
\end{equation}
and $P_\epsilon$ has $20$ conifold points instead of $24$; there are $8$ at $\{Q=x_3=0\}\cap S_\infty$,
$8$ at $\{Q=x_4=0\}\cap S_\infty$, and $4$ at $\{x_3=x_4=0\}\cap S_\infty$, so the deformation
smoothed four conifold points.

The reason for deforming $P$ in this way -- that is, ``picking'' out
the $x_1$ and $x_2$ coordinates to put in $Q$ rather than some
other set -- is that the four conifold points that were lost were
those resolved by the third of the four blow-ups,
i.e., $\pi_2$. This means that we can blow up $P_\epsilon$ in the
same order, to give new deformed varieties as
\begin{equation}
P_{43,\epsilon}\xrightarrow{\pi_3}P_{4,\epsilon}\xrightarrow{\pi_4} P_\epsilon,
\end{equation}
with $P_{43,\epsilon}$ smooth. This is because the $12$ conifold points $\{Q=x_4=0\}\cup\{x_3=x_4=0\}$ are resolved
by $\pi_4$ and the $8$ conifold points $\{Q=x_3=0\}$ are resolved by $\pi_3$. Then the
map
\begin{equation}
P_{43}\xrightarrow{\pi_3}
P_{4}\xrightarrow{\pi_4}
P\xrightarrow{\pi_\epsilon}
P_{\epsilon}\xrightarrow{\pi_4^{-1}}
P_{4,\epsilon}\xrightarrow{\pi_3^{-1}}
P_{43,\epsilon}
\end{equation}
means that there is a deformation of $P_{43}$, which has four conifold points, to $P_{43,\epsilon}$
which is smooth.

In summary, we have a map $Z\rightarrow P_{43}$ which blows down four
rigid rational curves of class $\gamma_1-\gamma_2$, yielding four
conifold points. There is another small resolution of $Z'\rightarrow
P_{43}$ where $Z$ and $Z'$ are related via a flop transition, where
the curves of class $\gamma_1-\gamma_2$ are flopped. There is also a
deformation $P_{43}\rightarrow P_{43,\epsilon}$ which deform the four
conifold points in $P_{43}$, so that $Z$ and $Z'$
are related to $P_{43,\epsilon}$ via a conifold transition.

\sskip\sskip

Now we must demonstrate how this transition in the building block affects the
topology of the $G_2$ manifold $X$ and discuss how one might induce this transition
via movement in $G_2$ moduli.

First, we determine topology of the $G_2$ manifolds that would be
produced in the transition, should such transitions exist in moduli
space. Denote the $G_2$ manifolds obtained via the other small
resolution and deformation of $P_{43}$ as $X_s$ and $X_d$,
respectively. Since the building blocks of the two small resolutions
of $P_{43}$ are related by a flop of four rigid holomorphic curves
away from the neck, $K$ does not change and therefore
$b_2(X)=b_2(X_s)$. Since the three-cycles that appear in the two small
resolutions are in one to one correspondence with the appearance of
two-cycles, which are the same in number, we also have
$b_3(X)=b_3(X_s)$. Now consider the deformation. Since there are four
conifold points in $P_{43}$, the deformation to the smooth manifold
$P_{43,\epsilon}$ produces three-spheres which are expected to be\footnote{Technically,
even in the Calabi-Yau case, this is only known for non-compact examples.
 The existence of special Lagrangian representatives of the
new class associated to the three-spheres of deformation
is a common assumption in the literature that we also make. } special Lagrangian.
While we will say more in the physics discussion momentarily,
$\operatorname{dim}(K_s)=2$ where $K_s$ is $K$-lattice of the building block
associated to $P_{43,\epsilon}$. Therefore $b_2(X) = b_2(X_d)+1$.
However, though a three-sphere appears in the deformation, recall that
a three-cycle diffeomorphic to $S^2\times S^1$ vanishes in the
blow-down to $P_{43}$; therefore $b_3(X)=b_3(X_d)$. In summary, the
$G_2$ manifolds have Betti numbers related by
\begin{equation}
b_2(X)=b_2(X_s)=b_2(X_d)+1 \qquad \qquad b_3(X)=b_3(X_s)=b_3(X_d)
\label{eq:topology change}
\end{equation}
for the non-isolated flop and conifold transitions, respectively.

Now we argue that such topological transitions should actually exist via
movement in $G_2$ moduli. Kovalev's theorem guarantees the existence
of a torsion-free $G_2$ structure $\Phi$ that is a small deformation
of the natural $G_2$ structure $\Phi_{T,r}$ on the twisted connected
sum, and moreover $[\Phi]=[\Phi_{T,r}]$. Now, $H^3(X,\bZ)$ and
$H^2(X,\bZ)$ both contain $K_+ \oplus K_-$, and in fact recall from
section \ref{sec:TCS review} that for $\alpha_\pm \in K_\pm$ we have
\begin{equation}
\alpha_\pm \ne 0 \in H^2(X,\bZ) \qquad \text{and} \qquad \alpha_\pm\wedge d\theta_\mp \ne 0 \in H^3(X,\bZ).
\end{equation}
Therefore choosing an integral basis of $H^3(X,\bZ)$ in which to expand $\Phi$ we see
\begin{equation}
\Phi = \sum_{i=1}^{rk(K_+)} \phi_i \,\, \alpha_+^i \wedge d\theta_- + \dots
\end{equation}
for the $G_2$-form. One might also think of this suggestively as
\begin{equation}
\Phi = (\sum_{i=1}^{rk(K_+)} \phi_i \,\, \alpha_+^i) \wedge d\theta_- + \dots
\end{equation}
and we wish to integrate this three-form  over one of these associative threefolds
diffeomorphic to $S^2\times S_-^1$.  Now if we integrate the nearby form
$\Phi_{T,r}$, the integral $\int_{S^1_-} \,
d\theta_- = 2\pi R_-$ factors out and does not depend on moduli. We
anticipate that passing from $\Phi_{T,r}$ to $\Phi$ changes this behavior
somewhat, but the change should not be large.  In particular, if we
are close to the point in the moduli space of the building block where
the singularity appears, and if $R_-$ is large, we would expect small 
corrections so that the integral of $S_-^1$ remains positive even when
the transition point in moduli is reached.\footnote{Of course, this
statement should be mathematically proven, if possible!} If so, then if we
vary $\phi_i$ such that a rigid associative vanishes,
the $S^1$ stays
at finite volume and therefore the $S^2$ must vanish. So we expect to be 
able to
control the two-cycles via their relation to these calibrated
three-cycles. This is our argument in favor of
 the existence of the non-isolated $G_2$
flop.\footnote{CHNP point out that a more complete mathematical treatment
of the non-isolated $G_2$ flop and $G_2$ conifold transitions would involve proving that
the singular space has an appropriate metric, which is a limit of metrics
on nearby nonsingular spaces.  This same point can be
made about the flop \cite{Witten:1993yc,mult, catp}
and conifold \cite{bhole} transitions for Calabi--Yau threefolds,
where the metric is known for local models \cite{CdO} but not for
global models.  (Metrics are also known for local models of the
isolated $G_2$ flop and conifold transitions \cite{Brandhuber:2001yi,Cvetic:2001zx,Cvetic:2001ih}.)}

To argue for the existence of the non-isolated
$G_2$ conifold, we must also be able
to control the three-cycles produced in the deformation of $P_{43}$ in
$G_2$ moduli. As discussed above, in the building block the deformation
is expected to produce four special Lagrangian three-spheres, which of course have
$b_1=0$. Since the three-spheres are not cycles in $S_s\times S^1$ but are in
$H_3(V_s,\bZ)$, then by the
long exact sequence in relative homology
\begin{equation}
\dots \to H_3(S_s\times S^1,\bZ)\to H_3(V_s,\bZ) \to H_3(V_s,S_s\times S^1) \to H_2(S_s\times S^1,\bZ) \to \dots
\end{equation}
they are non-trivial in $H_3(V_s,\bZ)$. So these three-spheres are
expected to
satisfy the conditions of the theorem of \cite{Corti:2012kd} discussed in section
\ref{sec:TCS review}, and therefore a small deformation of any one of
them is expected to give
an associative in the $G_2$ manifold $X_s$. Such associatives can be used to control
the deformation of the $G_2$ conifold in $G_2$ moduli.

In summary, we have argued for the existence of non-isolated flop and conifold
transitions beginning with the TCS $G_2$ manifold $X$. The topology
change from $X$ to the other small resolution $X_s$ and the
deformation $X_d$ is given in (\ref{eq:topology change}). We have
explained in each case how  movement in $G_2$ moduli could cause
associative submanifolds in the $G_2$ manifolds $X$, $X_s$, and $X_d$
to vanish or grow in such a way that flop and/or conifold transitions
may be induced by the one in the building block.

\subsection{Physics of the Topology Change}
\label{sec:physics of topology change}

We now discuss the physics of M-theory on the branches of $G_2$ moduli
related by topology change, using $X$, $X_c$, $X_s$ and $X_d$ for the original
manifold, the singular limit with circles of conifolds, the other $G_2$ small resolution,
and the $G_2$ deformation, respectively.

The ``$G_2$ blow-down'' $X\to X_c$ is a limit in which the volumes of
the four rigid
associatives associated to rigid curves of class $\gamma_1-\gamma_2$
vanish; accordingly, the massive chiral multiplet made of $\Psi_{14}^i$
and $\ov \Psi_{14}^i$ becomes massless in the limit. At large volume
there are instanton corrections as given in (\ref{eq:explicit
  instanton corrections}), and as the limit is approached there may be
instanton corrections that are subleading at large volume that become
important.

M-theory on $X_c$ therefore has massless particles charged under $U(1)^3$
with charges
\begin{center}
  \begin{tabular}{c|c|c|c|}
    & $Q_1$ & $Q_2$ & $Q_3$ \\ \hline
 $\Psi^k_{12}$ & $-2$ & $-1$  & $-1$ \\ \hline
 $\ov\Psi^k_{12}$ & $2$ & $1$  & $1$ \\ \hline
  \end{tabular}
\end{center}
in addition to massive particles with the same charge as the massive particles
for M-theory on $X$ that did not become massless.

M-theory on $X_s$, obtained via a $G_2$ small resolution from $X_c$
via associative threefolds as discussed in the last section, has
$U(1)^3$ gauge symmetry and a spectrum of charged particles identical
to that of M-theory on $X$; though the curve classes flopped as
$\gamma_1-\gamma_2 \mapsto \gamma_2-\gamma_1$, anti M2-branes on
curves of the latter class in $X_s$ have the same charges as M2-branes
on curves of the former class in $X$, so the overall set of particle
charges remains the same. M-theory on $X_s$ also exhibits a
non-perturbative superpotential generated by membrane instantons, but
the rigid associative associated to the flopped rigid holomorphic
curve has a relative sign, so that the non-perturbative superpotential
is identical to (\ref{eq:explicit instanton corrections}) except for
the replacement $\Phi_1\mapsto -\Phi_1$.

\sskip \sskip M-theory on $X_d$ is slightly more complicated, and so
we will devote a few paragraphs to some details that we have not yet
discussed. Since
$b_2(X_d) = b_2(X)-1=2$, we know that that M-theory on $X_d$ exhibits
$U(1)^2$ gauge symmetry rather than the $U(1)^3$ of M-theory on $X_c$.

First we give a simple field theoretic argument for what must be true of
M-theory on $X_d$. Note that since the only massless charged fields at
the singular point $X_c$ have charges $\pm(-2,-1,-1)$ under $U(1)^3$,
these must be the fields which spontaneously break one of the $U(1)$
symmetries. These fields are uncharged under the combinations $\tilde
Q_1 \equiv Q_1-2Q_2$ and $\tilde Q_2\equiv Q_1-2Q_3$, and therefore
these must be the two $U(1)$ symmetries which exist for M-theory on
$X_d$ (up to redefinition).  On $X$ these $U(1)$'s have generators
$E_1-E_2-2(E_1-E_3)=2E_3-E_2-E_1\equiv 2E_3 - E$ and similarly
$2E_4-E_2-E_1=2E_4-E$, respectively; we note that both generators have a
common term $E$.

This field theoretic argument must match the topology of $X_d$, since the latter
determines the particle charges. How does one
see this? Since the intersection theory was originally determined by
blowing up along curves rather than divisors, we will do the same here.
If we perform the deformation of the pencil in $\bP^3$ as discussed
\begin{equation}
|S_0,S_\infty| \to |S_{0,\epsilon},S_\infty|
\end{equation}
then its base locus is now a union of three curves instead of four;
two of them are again $C_3$ and $C_4$, but $C_1$ and $C_2$ have been
replaced by $C_Q\equiv \{S_\infty = Q=0\}$. So the base locus of
$|S_{0,\epsilon},S_\infty|$ is the union of $C_3, C_4,$ and $C_Q$ with
classes $[C_3]=[C_4]=H^2$ and $[C_Q]=2H^2$. Blow up along the curves
of the base locus sequentially in the order $C_Q,C_3,C_4$. Now we have
three exceptional divisors, $E_Q, E_3$ and $E_4$, respectively, which
restrict to curves of class $H$, $H$, and $2H$ in $S$,
respectively. The $K$-lattice of the deformation $K_d$ is therefore
generated by $2E_3-E_Q$ and $2E_4-E_Q$, and therefore these are the
generators of $U(1)^2$ for M-theory on $X_d$; note that they look
identical to what the field theoretical answer required, but let us
compute particle charges as a rigorous check. Letting $\gamma_Q$,
$\gamma_3$ and $\gamma_4$ be the class of the generic fiber of $E_Q$,
$E_3$, and $E_4$, the particles come from M2-branes wrapped on $8$ rigid
holomorphic curves of class $\gamma_Q-\gamma_3$, $8$ of class
$\gamma_Q-\gamma_3$, and $4$ of class $\gamma_3 - \gamma_4$. The
intersection theory is computed as before with the result $E_Q\cdot
\gamma_Q = E_3 \cdot \gamma_3 = E_4\cdot \gamma_4=-1$, as are the particle
charges via the intersections of the particle curves with the $U(1)$ generating
divisors within the building block. For example, under the $U(1)$ of $2E_3-E_Q$
the particles on curves of class $\gamma_Q-\gamma_3$ have charge $3$.

A short computation shows that the topological calculation of particle charges
matches the field theoretic expectation from the previous paragraph. The result is that
the generators of $K_d$ (and thus of $U(1)$'s) $2E_3-E_Q$ and $2E_4-E_Q$  correspond precisely
to $\tilde Q_1$ and $\tilde Q_2$. M-theory on $X_d$ exhibits massive particles with
charge
\begin{center}
  \begin{tabular}{c|c|c|}
    & $\tilde Q_1$ & $\tilde Q_2$ \\ \hline
 $\Psi^j_{Q3}$ & $3$ & $1$ \\ \hline
 $\Psi^j_{Q4}$ & $1$ & $3$ \\ \hline
 $\Psi^k_{34}$ & $-2$ & $2$ \\ \hline
  \end{tabular}
\end{center}
and their conjugates, where $j=1\dots 8$ and $k=1\dots 4$.

It is satisfying that the field theory prediction of
particle charges after $U(1)^3\to U(1)^2$ symmetry breaking matched
the topological computation after the non-isolated $G_2$ conifold
transition.

\section{The Landscape of M-theory on $G_2$ Manifolds}
\label{sec:landscape}

Having reviewed the twisted connected sum construction and studied a rich
example with many interesting physical features, in this section
we would like to study some aspects of the associated landscape of (abelian)
four-dimensional $\cN=1$ compactifications. Some of the general statements
about the physics of $G_2$ compactifications that were introduced in the example
will be reintroduced for the sake of completeness.

\sskip
\sskip

What is a coarse measure of the size of the (known) abelian $G_2$
landscape and how has it grown in recent years?  We can be more
precise than we were in the introduction, since we have introduced the
TCS construction.  The earliest examples of compact $G_2$ manifolds
due to Joyce \cite{MR1424428} were relatively few in number.  The
original TCS examples \cite{KovalevTCS} utilized building blocks of
Fano type, and the number of such examples is determined in part by
the number of smooth Fano threefolds; these have been classified and
there are precisely $105$ deformation families. By contrast,
\cite{Corti:2012kd} also constructs $G_2$ manifolds using semi-Fano
building blocks, which utilize weak-Fano threefolds; there are at
least hundreds of thousands of deformation families of smooth
weak-Fano threefolds. These give rise to \cite{Corti:2012kd} at least
$50$ million matching pairs, arising only from ACyl Calabi-Yau
threefolds from semi-Fano building blocks of rank at most two or from
toric semi-Fano threefolds; given the limited nature of this search,
many more can probably be obtained by considering higher rank building
blocks. These $50$ million matching pairs from semi-Fano building
blocks each give rise to a TCS $G_2$ manifold and, therefore, a
four-dimensional M-theory vacuum, some of which may be equivalent in
cases where the same $G_2$ manifold arises from different building
blocks. It is noteworthy that the number of $G_2$ building blocks is
within a factor of ten of the number of Kreuzer-Skarke ``Calabi-Yau
building blocks,'' i.e., the $500$ million reflexive four-dimensional
polytopes.  

The number of known abelian $G_2$ compactifications is already quite large
and via systematic application of existing construction techniques it
will likely continue to grow in the coming years. Given this large
number of examples and the already existing evidence for topology
change, it seems reasonable to wonder whether $G_2$ moduli space is
connected, as Reid has conjectured \cite{MR909231} for
Calabi-Yau threefolds. Such a property would strengthen the meaning of
the abelian $G_2$ landscape, as the associated vacua would form a connected
moduli space of a single theory.

\sskip \sskip Finally, though it will not be critical in the
following, it is worth mentioning that TCS $G_2$ manifolds often have
large numbers of moduli. For example, Table 5 of \cite{Corti:2012kd}
lists a few dozen $G_2$ manifolds with $47\leq b_3(X)\leq 155$, taking
many different values in this range; models with only a handful of
moduli are scarce.

\subsection{Higgs Branches, Coulomb Branches, and Gauge Enhancement}
\label{sec:Higgs}

In \cite{SingularLimits} we will study a number of different ways in
which one might take a singular limits of a $G_2$ compactification in
order to obtain non-abelian gauge enhancement or massless charged
matter in the theory.\footnote{For some earlier work on non-abelian gauge symmetries
in $G_2$ compactifications, see \cite{Acharya:1998pm,Acharya:2000gb} or the review \cite{Acharya:2004qe}. For the relationship between these constructions
and chiral type IIa constructions with intersecting D6-branes, see \cite{Cvetic:2001kk,Cvetic:2001nr,Cvetic:2001tj}.}  If an M-theory compactification on a $G_2$
manifold $X$ admits a limit in which non-abelian gauge enhancement
occurs, then a natural question is whether the the vacuum is on a
Higgs branch or a Coulomb branch.

In the bulk of this section we will not look in detail at Higgsing from a
non-abelian theory, instead speaking of ``Higgs branches'' and ``Coulomb
branches'' loosely according to the value of
$b_2(X)$.  For now, let us be slightly
more precise. Suppose there existed a singular limit of $X$ which
realizes a gauge sector with gauge group $G\times U(1)^k$, where $G$
is the nonabelian part (with finite center). If
smoothing the manifold back to $X$ (or another member in the same
family of $G_2$ manifolds as $X$) Higgses this theory in a standard
way, then an upper bound on the number of $U(1)$'s is set by the
dimension of the maximal torus of the gauge theory on the singular
space; that is
\begin{equation}
b_2(X) \leq rk(G) + k.
\end{equation}
Certainly if a gauge enhanced singular limit exists and $b_2(X)=0$ then
the vacuum obtained from M-theory on $X$ is on a Higgs branch;
conversely $b_2(X)\ne 0$ is necessary for this vacuum to be on a
Coulomb branch. That said, if $G$ is finite but $k\ne 0$
then $b_2(X)=k$ and the terminology is slightly ambiguous since there
still are long range forces but none of them arise from Cartan
$U(1)$'s of a non-abelian $G$.

Despite these caveats, we will loosely call these vacua with
$b_2(X)=0$ Higgs branches and vacua with $b_2(X)\ne 0$ Coulomb
branches. Vacua of the latter type are particularly useful since, for
example, particle charges can be computed, and if they are the charges of
massive W-bosons of a spontaneously broken gauge theory then the
charges are intimately related to gauge enhancement in a singular
limit.

\subsubsection*{Higgs Branches: Their Prevalence and Drawbacks}

It turns out that \emph{nearly all} of the known examples are on Higgs
branches; i.e., they have $b_2(X)=0$.  This follows from topological
properties of the building blocks used to construct most TCS $G_2$
manifolds.  Since for a semi-Fano building block the map
$\rho:H^2(V)\rightarrow H^2(S)$ is injective, $K=0$; then for any TCS
$G_2$ manifold built out of two semi-Fano building blocks
\begin{equation}
H^2(X,\bZ) = N_+ \cap N_-.
\end{equation}
However in order construct a TCS $G_2$ manifold from the building
blocks, one must also solve the matching problem; i.e., there must
exist a \KM\ $r:S_+\rightarrow S_-$. This problem
is much easier to solve if $N_+ \cap N_-=0,$ in which case we have
$H^2(X,\bZ)=0$.  M-theory on such an $X$ is on a Higgs branch
if a singular limit with non-abelian gauge symmetry exists.

We can be slightly more specific. A manifold $X$ is said to be
2-connected if $\pi_1(X)=\pi_2(X)=0$; then we also have
$H^1(X)=H^2(X)=0$.  A smooth 2-connected seven-manifold --- and
therefore a smooth 2-connected $G_2$ manifold --- is classified up to
almost-diffeomorphism\footnote{An almost-diffeomorphism is an invertible
map which is smooth except possibly at a finite number of points.
The classification given in \cite{MR0307258} was recently
sharpened to a diffeomorphism classification by introducing
additional invariants \cite{arXiv:1406.2226}.}
by the pair of non-negative integers
$(b_4(X),div\, p_1(X))$, where $div\, p_1(X)$ measures the
divisibility of the first Pontryagin class \cite{MR0307258}.
  It so happens that the
$50$ million matching pairs discussed in \cite{MR3109862,Corti:2012kd}
give rise to 2-connected $G_2$ manifolds. Therefore since $H^2(X)=0$
for all of these manifolds, any of the associated M-theory vacua are
on Higgs branches. Though (as we
saw in section \ref{sec:example}) there are known vacua with $b_2(X)\ne 0$, which
are on Coulomb branches if some of the associated $U(1)$'s embed into
a non-abelian group in a singular limit, essentially the entire known
$G_2$ landscape is comprised of Higgs branches.

\sskip

In practice, studying singular limits of $G_2$ compactifications that
exhibit massless charged matter or non-abelian gauge enhancement is
much more difficult when approaching from Higgs branches rather than
from Coulomb branches.  One physical reason is that the Higgs vacuum
does not exhibit any charged particles since the gauge symmetry is
completely broken; therefore there is no charge to ``measure'' (as we
did in section \ref{sec:example}) via the intersection theory of
non-trivial two-cycles and five-cycles in $X$. Moreover, it is
conceivable that progress could be made in $G_2$ Higgs vacua similar to
the recent progress in F-theory Higgs vacua. For example, in the
latter case it is known
\cite{Grassi:2013kha,Grassi:2014sda,Grassi:2014ffa} how to recover the
spectrum of massive W-bosons in a completely Higgsed theory via the study
of an elliptic fibration. It is possible that some $G_2$ manifolds
admit a similar elliptic fibration \cite{Vafa:1996xn,Grassi:2014ffa}. 

\subsubsection*{Coulomb Branches: Their Scarcity and Utility}

Though most TCS $G_2$ vacua are on Higgs branches, vacua on Coulomb
branches do exist. In fact, most of the original examples of Joyce
\cite{MR1424428} were Coulomb branches due to $H^2(X)$ being
non-trivial. Those examples were seven-manifolds with ADE
singularities, and upon smoothing to the $G_2$ manifold many expected
features occur. In \cite{SingularLimits} we will study the physics of
some Joyce manifolds, where the simplest cases involve moving from a
non-abelian theory $G$ to $U(1)^{rk(G)}$ via adjoint breaking. We will
also find a number of topological defects appeared in the Joyce
manifolds, for example the 't Hooft-Polyakov monopoles characteristic
of symmetry breaking to Coulomb branches.

To obtain a TCS $G_2$ vacuum on a Coulomb branch it is necessary to study
examples with non-trivial $H^2(X)$; since
\begin{equation}
H^2(X,\bZ) = (N_+ \cap N_-)\oplus K_+ \oplus K_-
\end{equation}
some combination of $K_\pm$ and $N_+\cap N_-$ must be non-trivial. One
option is to use so-called non-perpendicular orthogonal gluing, in
which case $N_+\cap N_-$ is non-trivial; the drawback of this option
is that it can be difficult to find a \KM. The
other option is to use a building block with $K\ne 0$; but then the
building block cannot be taken from the large collection of semi-Fano
blocks which all have $K=0$.

In \cite{Corti:2012kd} two possible methods were suggested for
constructing building blocks with $K \ne 0$, and thus $G_2$
compactifications on Coulomb branches. The first is to construct new
building blocks obtained by blowing up a non-generic anticanonical
(AC) pencil in a toric semi-Fano 3-fold, which \emph{is not} a
``semi-Fano building block'' since the latter assumes a blow-up of a
generic AC pencil. \cite{MR3109862,Corti:2012kd} identified some
examples of this type and computed properties of the associated
lattices $K$. Given that the basic object of this approach is a toric
variety associated to a three-dimensional reflexive polytope, it would
be interesting to study whether the ``non-generic AC'' pencil
construction of building blocks with $K\ne 0$ can be systematized, and
if so what the associated physics is on the Coulomb branch.  The other
suggestion of \cite{Corti:2012kd} for constructing building blocks
with $K\ne 0$ is to use one of the $74$ non-symplectic type building
blocks introduced in \cite{KovalevLee}. In section \ref{sec:example}
we studied an example of the former type and saw that many physical
effects can be computed.

Though they are relatively scarce (for technical reasons) in the class
of known TCS $G_2$ manifold examples, those that describe M-theory
vacua on Coulomb branches, or more generally vacua with $U(1)$
symmetries and massive charged particles are practically useful and
physically interesting. Suppose M-theory on $X$ yielded a Coulomb
branch vacuum. Then massive W-bosons from M2-branes wrapped on two-cycles are part of the
charged particle spectrum of the theory and it is clear what to look
for: limits in $G_2$ moduli space in which those two-cycles go to zero
volume.

Though there is no calibration form for two-cycles in a $G_2$
manifold, one can imagine cases where two-cycles are contained in an
associative threefold or coassociative fourfold; the natural question
in such a case is whether some degenerations (zero volume limits) of
the associative or coassociative submanifolds give rise to collapses
of the two-cycles they contain. This can be one handle on obtaining
non-abelian gauge enhancement or massless charged particles, as we
saw explicitly in the example of section \ref{sec:example}. There two-cycle
volumes were directly controllable via associative threefolds; we will
discuss this idea further in our work \cite{SingularLimits}.

\subsection{Membrane Instantons, $G_2$ Transitions, and Fluxes}
\label{sec:instantons}

The results of \cite{Corti:2012kd} have a number of other implications
for the physics of M-theory compactifications on TCS $G_2$ manifolds,
as we will discuss in this section. 

\sskip
\sskip
\noindent \textbf{\emph{Instantons and Rigid Associatives}}

Instantons effects arising from wrapped branes and strings can generate non-perturbative
corrections to the scalar potential that play an important role in moduli stabilization.
In M-theory compactifications these may arise from wrapped M2-brane or M5-brane
instanton corrections to the superpotential. While M5-brane instantons
play a major role in M-theory compactifications on  Calabi-Yau fourfolds
to three dimensions \cite{Witten:1996bn}, for example by providing effects that
lift Coulomb branches, these corrections do not exist in $G_2$ compactifications
since seven-manifolds with holonomy precisely $G_2$ have $b_6(X)=0$; that is,
there are no cycles on which to wrap M5-brane instantons.

In contrast, in $G_2$ compactifications an M2-brane instanton may
generate a superpotential correction if it is wrapped on a rigid
supersymmetric (i.e., associative) three-cycle
\cite{Harvey:1999as}. These instanton corrections to the superpotential
$W$ take the heuristic
form
\begin{equation}
 A e^{-\Phi}
\end{equation}
for the $G_2$ modulus $\Phi$ associated to the rigid associative
three-cycle wrapped by the instanton.  While it is not yet possible to
make complete statements about the structure of the instanton
prefactor $A$ due to the absence of a microscopic description for
instanton zero modes, in the analogous D-brane instanton cases in
F-theory and/or type IIa the prefactor $A$ may contain chiral matter
insertions or an intricate geometric moduli dependence (see
e.g. \cite{Blumenhagen:2006xt} and \cite{Cvetic:2012ts,Kerstan:2012cy}).
Note that the absolute value of the prefactor should be given by the $\eta$
function of an appropriate Dirac operator \cite{Harvey:1999as}.

\sskip
\sskip

Twisted connected sum $G_2$ compactifications are currently the only
$G_2$ compactifications where one may concretely study instanton
corrections to the superpotential, since the first compact rigid
associative cycles in a $G_2$ manifold were constructed in
\cite{Corti:2012kd}, and this construction is specific to twisted
connected sum $G_2$ manifolds. The relevant theorem is that if $C$
is a rigid holomorphic curve in $V$, then a small deformation of
$S^1\times C$ in $X$ is a rigid associative. An M2-brane on this rigid
associative corrects the superpotential in models where interactions lift
 Wilson line modulini; see the footnote in section \ref{sec:massive charged particles on X}.

While this gives a method for identifying compact rigid
associatives in TCS $G_2$ manifolds, there may exist others rigid
associatives that are of a different type. We emphasize this point
because it means that while current techniques allow for the
identification of some instanton corrections, it is not yet possible
to say whether these are all of the corrections, or even the leading
corrections. Thus, explicit $G_2$ moduli stabilization via instantons is still
out of reach.

That issue aside, how many instanton corrections exist in known
examples?  For some of the examples in \cite{Corti:2012kd} the number
$a_0$ of rigid associatives associated to rigid holomorphic curves in
the building blocks was computed, with $a_0$ ranging from $0$ to $66$,
taking a variety of values in between. In the example of M-theory on
$X$ that we studied in section \ref{sec:example}, $a_0=24$ due to the
existence of $24$ rigid holomorphic curves in one of the building
blocks; after the studied non-isolated $G_2$ flop (conifold) transition the
geometry had $a_0=24$ ($a_0=20$).  Note that some rigid associatives
may be in the same homology class (as in the example), in which case
there is an associated multiplicity factor in front of the instanton
correction, which should be thought of as the M-theory on $G_2$ analog
of Gromov-Witten invariant prefactors of worldsheet instantons on Calabi-Yau
threefolds;
recall, for example, that there are $2875$ lines in the quintic which
give rise to instanton corrections from string worldsheets, though all
in the same homology class.

If some number of these rigid associatives are in different homology classes,
however, the superpotential takes the form of a racetrack or a generalized
racetrack with multiple terms, i.e.,
\begin{equation}
W_\text{inst} =  \sum_i A_i e^{-\Phi_i}
\end{equation}
where $\Phi_I$ is the chiral multiplet modulus associated to the rigid associative via
Kaluza-Klein reduction.
More specifically in the example we studied the superpotential took the form
\begin{equation}
W = 4(A_{1}e^{-\Phi_1}+A_{2}e^{-\Phi_2}+A_{3}e^{-\Phi_3}+A_{4}e^{\Phi_1-\Phi_2}+A_{5}e^{\Phi_1-\Phi_3}+A_{6}e^{\Phi_2-\Phi_3}) + \dots
\end{equation}
which is a six-term generalized racetrack.  In a singular limit of $X$ there may be
additional terms of this structure in the superpotential due to a
confining hidden gauge sector; see the studies
\cite{Acharya:2006ia,Acharya:2007rc} which utilize hidden sectors.
Based on this evidence, it seems that racetracks or generalized
racetracks occur frequently.

\sskip
\sskip
\noindent \textbf{\emph{$G_2$ Transitions}}

Given the existence of flop and conifold transitions for string
compactifications on Calabi-Yau threefolds, it is natural to wonder
about the possibility of topology changing transitions in $G_2$
compactifications of M-theory. The existence of such a transition
would require two topologically distinct families of $G_2$ manifolds
which give the same singular space in some limit of their respective
moduli spaces. The transition would occur by taking the limit of one
of the families, and then passing to the other family via the
intermediate singular space. This is the natural analog for $G_2$
manifolds of a flop or conifold transition, as already explored in 
an example in section \ref{sec:example}.

In general there are still difficulties with establishing the
existence of $G_2$ transitions for TCS $G_2$ manifolds, partly
because of difficulties in controlling the sizes of the corrections
to $\Phi_{T,r}$ as moduli are varied,
but there is an interesting and natural possibility for
realizing these transitions given that the building blocks are composed of
algebraic threefolds. Namely, if the algebraic threefold of the
building block can \emph{itself} undergo a transition and on both
sides of the $G_2$ transition the TCS construction can be used to
construct topologically distinct $G_2$ manifolds, then one might study
whether the associated transition between $G_2$ manifolds exists via
movement in $G_2$ moduli. This was precisely what we did in section
\ref{sec:example}, utilizing the fact that two-cycle volumes should be
controlled via related associative submanifolds. We found that there
should be (non-isolated) $G_2$ flop and conifold transitions 
related to flop and conifold transitions in a building block.

In \cite{Corti:2012kd} a number of interesting general observations
were made in about $G_2$ transitions which are induced by conifold transitions
in the building blocks.  Suppose there is a conifold transition
$F\rightarrow \tilde X \rightarrow Y$ between a smooth Fano $F$ and a
smooth semi-Fano $Y$ via an intermediate singular threefold $\tilde
X$. Suppose further that one is able to use the associated building
blocks $(Z_Y,S_Y)$ and $(Z_F,S_F)$ to construct TCS $G_2$ manifolds
$X_Y$ and $X_F$. Then it is natural to wonder whether there is a $G_2$
transition from $X_F$ to $X_Y$ associated to the threefold transition
from $F$ to $Y$. In \cite{Corti:2012kd} it is observed that
\begin{itemize}
\item[1)] $b_2(Y)>b_2(F)$
\item[2)] $b_3(Y)\leq b_3(F)$, and in fact it typically is a strict inequality
\item[3)] $Y$, but not $F$, contains compact rigid rational curves which do not intersect smooth
  anticanonical divisors and give rise to compact rigid rational curves in
  the associated ACyl $CY_3$ $Z_Y\setminus S_Y$.
\end{itemize}
At the level of constructing associated $G_2$ manifolds, the authors
note that 1) implies that solving the matching problem for building
blocks constructed from $Y$ is more difficult than for those constructed
from $F$; that 2) implies that $b_3(X_Y)\leq b_3(X_F)$, i.e., the number
of $G_2$ moduli often changes; and that 3) implies that
the rigid rational curves of $Y$ give rise to compact rigid
associatives in $X_Y$ which do not exist in $X_F$.

We would like to note that each of these observations has interesting
physical consequences in the associated M-theory compactifications.
The associated physical statements are:
\begin{itemize}
\item[1)] Since changing $b_2$ of the building blocks does not necessarily
  change $b_2$ of the associated $G_2$ manifolds, such a $G_2$ transition
  could in principle be a Higgs-Higgs transition or a Higgs-Coulomb transition (if
  gauge symmetry exists on the singular space at all), whereas for a conifold
  transition in string theory it is a Higgs-Coulomb transition.
\item[2)] A change in the number of moduli has implications for moduli
  stabilization, but there are also corresponding instantons (which do
  not necessarily correct $W$), domain walls, and axion strings that
  appear in the compactification on $F$ due to wrapped M-branes.
\item[3)] There are instanton corrections to the superpotential for
  M-theory on $Y$ that do not exist for M-theory on $F$; this is
  similar to behavior elsewhere in the landscape, for example in the
  Higgs-Coulomb transition that may arise for three-dimensional
  M-theory compactifications on elliptically fibered Calabi-Yau
  fourfolds.
\end{itemize}
Again, we emphasize that these are physical statements following from
the topology of potential TCS $G_2$ transitions which may be induced
by transitions in the algebraic building blocks; the topological
statements are true, but there may not exist $G_2$ metrics
throughout the proposed transition. It would be interesting to study
whether they exist in broad classes of examples. 

\sskip
\sskip
\noindent \textbf{\emph{Flux and Fluxless Compactifications}}

In M-theory compactifications it is possible to turn on four-form flux
$G_4$. Consider M-theory on a manifold $X$. For the theory to be well-defined, the flux must satisfy the quantization
condition \cite{Witten:1996md}
\begin{equation}
\left[\frac{G_4}{2\pi}\right] - \frac{p_1(X)}{4} \in H^4(X,\bZ)
\end{equation}
where $p_1(X)$ is the first Pontryagin class of $X$. This
flux quantization condition has an interesting corollary: since
this specific combination of four-forms must be integral, if $p_1(X)/4$
is not integral then choosing such a compactification manifold $X$
\emph{requires} $G_4 \ne 0$. This is a well-known phenomenon in F-theory,
where the elliptically fibered Calabi-Yau fourfold of the related M-theory
compactification sometimes requires that flux be turned on.

What about for M-theory compactifications on a $G_2$ manifold $X$? In
\cite{Corti:2012kd}, $p_1(X)$ was computed for the first time in terms
of data of the building blocks; thus, in concrete examples one can now
check whether $p_1(X)/4$ is integral. While the precise
knowledge of $p_1(X)$ is convenient, it is not necessary to answer the
question of whether flux must turned on, since it is known from
\cite{Beasley:2002db} that $p_1(X)/4$ is integral and thus one can
always consistently choose to set $G_4=0$ in any $G_2$
compactification. However, if $G_4\ne 0$ there is a perturbative
flux superpotential and moduli stabilization is qualitatively different;
one may also argue that this is more generic.

\sskip
\sskip
\noindent \textbf{\emph{Common Model-Building Assumptions in Light of TCS $G_2$ Manifolds}}

In studying the landscape scenarios are often put forth for moduli
stabilization and supersymmetry breaking based on sound theoretical
arguments and calculations, but before large classes of examples
exist; once they do exist, though, it is interesting to re-evaluate the
scenarios.

A well-known example is the large number of type IIb flux vacua, where
this large number arises from a large number of possible integral
Ramond-Ramond fluxes that may be chosen to stabilize the complex
structure of the Calabi-Yau $X$. Though the general calculations are
sound, typically quoted flux vacuum counts (e.g. $10^{500}$) exist for
large $h^{2,1}(X)\gtrsim 100$, and integral fluxes have never been
constructed for Calabi-Yau manifolds with such large Hodge numbers for
reasons of computational complexity. If this obstacle were removed, it
would be nice to have an explicit example which confirms the
assumptions and results of the proposed scenario.

Similarly, scenarios have been proposed for moduli stabilization and
supersymmetry breaking (as well as phenomenology) in $G_2$
compactifications of M-theory.  For example, in one scenario known at
the $G_2$-MSSM (see e.g. the review \cite{Acharya:2012tw}) at least
three important assumptions are made:
\begin{itemize}
\item[1)] The M-theory compactification is fluxless, i.e., $G_4$ is cohomologically trivial.
\item[2)] The primary source of moduli stabilization and supersymmetry
  breaking is from a strongly coupled hidden sector, which generates
  a non-perturbative superpotential $w_{np}$ containing terms of the form
  \begin{equation}
     A e^{-n_i \Phi_i} \qquad \qquad n_i \in \bZ
  \end{equation}
  where $\Phi_i$ are the metric moduli of the $G_2$ compactification
  and $A$ is determined by dimensional transmutation of the
  confining gauge theory.
  If this term drives moduli stabilization, membrane
  instanton corrections to the superpotential must be subleading.
\item[3)] The visible sector is an $SU(5)$ GUT broken to the MSSM
  via Wilson lines.
\end{itemize}
We would like to discuss some of these assumptions in light of the existence
of TCS $G_2$ compactifications and the associated physics discussed
in this section. We will address each in turn.

\sskip

The first assumption is always possible, since (as discussed) the flux
quantization condition never forces the introduction of $G_4$-flux in
a $G_2$ compactification of M-theory \cite{Beasley:2002db}, but
setting $G_4=0$ is also a non-generic choice, since it is
choosing the origin out of an entire vector space (the non-torsional
part of $H^4(X,\bZ)$). Interestingly, the absence of a flux superpotential
--- and therefore the choice $G_4=0$ ---
is critical in the moduli stabilization scenario of
\cite{Acharya:2012tw}. It would interesting to understand the extent
to which fluxes might alter the results of \cite{Acharya:2012tw}, or
whether the existence of de Sitter vacua depends in important ways on
the choice of flux or fluxless compactifications; the latter
dependence is plausible due to the fundamentally different structure
of the scalar potential in the two cases.

The second assumption is the one deserving the most scrutiny in
light of the recent progress. Since it has now been shown that
examples often exhibit many instanton corrections to the
superpotential and in the only explicitly computed example we found
the intricate form (\ref{eq:explicit instanton corrections}), it is
reasonable to expect that, at least in some cases, these effects will
compete with the non-perturbative superpotential of the confining
hidden sector utilized in \cite{Acharya:2012tw}. In a number of
examples of \cite{Corti:2012kd} there are over $40$ cycles which
support M2-brane instanton corrections to the superpotential. Though
(as discussed) these may be in the same homology class and thus generate
an exponentially suppressed correction in the same $G_2$ modulus,
\emph{at least one} instanton generated superpotential term exists in all of these
compactifications, and perhaps more if the rigid associatives are
homologically distinct; the example we studied is an existence proof
of the latter possibility. It would be interesting to understand how
the scenario \cite{Acharya:2012tw} changes when taking into account
instanton corrections; if it is a single new term it could, together
with the confining contribution, give a standard racetrack, whereas if
there are multiple distinct instanton corrections it would be a
generalized racetrack with many exponentially suppressed terms.

Not enough is known about singular limits of TCS $G_2$ manifolds to
evaluate the third assumption formally, beyond the typical arguments
made from heterotic / M-theory duality, since relatively little is
known about singular limits of compact $G_2$ manifolds as we will
discuss in \cite{SingularLimits}. Phenomenologically, it is an
assumption.

\section{Conclusions}
\label{sec:conclusions}

In this paper we have studied M-theory compactifications to four
dimensions on $G_2$ manifolds constructed via twisted connected
sum. There are now perhaps fifty million examples.

We have shown that recent topological progress \cite{Corti:2012kd} in
TCS $G_2$ manifolds now allows for interesting physical quantities to
be computed in the associated M-theory vacua on a TCS $G_2$ manifold
$X$. These include the $U(1)$ symmetries of the vacuum, the charges of
massive particles, the structure of some membrane instanton
corrections to the superpotential, spacetime topology change, and
spontaneous symmetry breaking in a $G_2$ conifold transition.

However, it is physically critical to understand singular limits of
these manifolds and their associated M-theory vacua. In our view, the
most important mathematical progress that would aid future physical
progress is to have a better understanding of singularities that
develop upon movement in $G_2$ moduli space, both in general and in
the twisted connected sum construction, since they are necessary for
realizing non-abelian gauge sectors or massless charged matter, and
therefore realistic vacua. In related work \cite{SingularLimits} we
will address a number of physical issues related to such degenerations
and will conjecture that the right approach will be to move to a wall
in a ``cone of effective associatives.''  In particular, as we will
discuss in \cite{SingularLimits} a critical physical issue for
understanding non-abelian gauge enhancement is to have some control
over intersections of two-cycles with five-cycles and limits in which
they degenerate. These degenerations are difficult to study since
there are no calibration forms for two-cycles; instead it would be
useful to have techniques to identify those cases in which a two-cycle
within an associative (coassociative) submanifold vanishes as the
associative (coassociative) itself vanishes. We saw such a phenomenon
in the example of section \ref{sec:example} due to a particular
factorization property which holds for certain cycles in TCS $G_2$
manifolds. While two-cycles are more difficult to study than three-
and four-cycles in a $G_2$ manifold, it would be important understand and
control them further since they determine the particle
physics of these M-theory vacua.

\sskip
It seems reasonable to hope that the singularities needed for non-abelian
gauge symmetry can eventually be engineered in the context of
the TCS construction,
either by finding a singular ACyl Calabi--Yau threefold with a singular
curve not extending to the boundary, or by extending the TCS construction
to allow the K3 surfaces along the neck to have rational double points.
We leave this to future work.

\sskip Our work is also a first step towards the explicit construction
of de Sitter vacua in fluxless $G_2$ compactifications of M-theory, as
we have computed the form of membrane instanton corrections to
the superpotential; see section \ref{sec:massless limits, topology change, and the Higgs mechanism} . These
instantons play an even more significant role for moduli stabilization
than their type IIb ED3-instanton counterparts, since in a smooth
$G_2$ compactification the fields which may be identified as moduli
are the metric and axion moduli which give $b_3(X)$ massless uncharged
chiral supermultiplets, and these are the fields that appear in
membrane instanton corrections to the superpotential.  Therefore,
membrane instanton corrections may in principle stabilize all moduli,
potentially giving rise to de Sitter vacua.

Physically, completing such a program requires having ``enough''
instantons to stabilize all moduli, and furthermore one must be able
to guarantee that these are the leading instanton
corrections. Mathematically, this requires the construction of
``enough'' associative submanifolds, ensuring that they are also
leading. While not completely precise, a rough way to think of the
``leading'' associatives is as follows. Let $T_i$ be an integral basis for
$H^3(X,\bZ)$. Then any rigid associative $M$ can be expanded in this basis as
$M=m_i\, T_i$. The leading instantons arise from instantons closer to the
origin where $m_i=0$ $\forall i$, so one algorithm would be to find all rigid
associatives in all homology classes in an appropriately sized box around the
origin.

While such vacua are
not realistic, giving rise to universes with axions and perhaps massive
charged particles and photons but no non-abelian gauge interactions,
they nevertheless would be de Sitter vacua. This may be the most
direct route to realizing de Sitter vacua in M-theory.
If the TCS construction can be extended to include singular limits
carrying non-abelian gauge fields, those de Sitter vacua could be
quite realistic.

\sskip\sskip

\noindent \textbf{Acknowledgments.} We thank Bobby Acharya, Allan Adams, Robert
Bryant, Mirjam Cveti{\v c}, Keshav Dasgupta, I\~naki Garc\'ia-Etxebarria, Antonella Grassi, Mark
Haskins, Gordy Kane, Nabil Iqbal, Albion Lawrence, Greg Moore,
Johannes Nordstr\"om, Joe Polchinski, Julius Shaneson, Washington
Taylor, and Scott Watson for useful conversations. We also thank the
Aspen Center for Physics and the Simons Center for Geometry and
Physics for hospitality during various stages of this project.  This
research was supported in part by the National Science Foundation
under grants PHY-1066293, PHY-1125915, and PHY-1307513.

\appendix


\bibliographystyle{JHEP}
\bibliography{newrefs}

\end{document}